\documentclass[traditabstract]{aa} 
\usepackage{txfonts}
\usepackage{graphicx}
\usepackage{longtable}
\usepackage{lscape}

\begin{document}

\title{Radial velocity variability and evolution of hot subdwarf stars}

\author{S.~Geier \inst{1}
   \and M.~Dorsch \inst{2}
   \and I.~Pelisoli \inst{3,1}
   \and N.~Reindl \inst{1}
   \and U.~Heber \inst{2}
   \and A.~Irrgang \inst{2}}
      
\offprints{S.\,Geier,\\ \email{sgeier@astro.physik.uni-potsdam.de}}

\institute{Institut f\"ur Physik und Astronomie, Universit\"at Potsdam, Haus 28, Karl-Liebknecht-Str. 24/25, D-14476 Potsdam-Golm, Germany
      \and Dr.~Karl~Remeis-Observatory \& ECAP, Astronomical Institute, Friedrich-Alexander University Erlangen-Nuremberg, Sternwartstr.~7, D-96049 Bamberg, Germany
      \and Department of Physics, University of Warwick, Coventry, CV4 7AL, UK}

\date{Received \ Accepted}

\abstract{Hot subdwarf stars represent a late and peculiar stage in the evolution of low-mass stars, because they are likely formed by close binary interactions. Here we performed a radial velocity (RV) variability study of a sample of 646 hot subdwarfs with multi-epoch radial velocities from Sloan Digital Sky Survey (SDSS) and Large Sky Area Multi-Object Fibre Spectroscopic Telescope (LAMOST) spectra. Atmospheric parameters and RVs were taken from the literature. For stars with archival spectra but without literature values, we determined the parameters by fitting model atmospheres. In addition, we redetermined the atmospheric parameters and RVs for all the He-enriched sdO/Bs. This large sample allowed us to study RV-variability as a function of the location in the $T_{\rm eff}-\log{g}$- and $T_{\rm eff}-\log{n({\rm He})/n({\rm H})}$ diagrams in a statistically significant way. As diagnostics we used the fraction of RV-variable stars and the distribution of the maximum RV variations $\Delta RV_{\rm max}$. Both indicators turned out to be quite inhomogeneous across the studied parameter ranges. A striking feature is the completely different behaviour of He-poor and He-rich hot subdwarfs. While the former have a high fraction of close binaries, almost no significant RV variations could be detected for the latter. This led us to the conclusion that there likely is no evolutionary connection between these subtypes.  Intermediate He-rich- and extreme He-rich sdOB/Os on the other hand are likely related. We conclude further that the vast majority of this population is formed via one or several binary merger channels. Hot subdwarfs with temperatures cooler than $\sim24\,000\,{\rm K}$ tend to show less and smaller RV-variations. These objects might constitute a new subpopulation of binaries with longer periods and late-type or compact companions. The RV-variability properties of the extreme horizontal branch (EHB) and corresponding post-EHB populations of the He-poor hot subdwarfs match and confirm the predicted evolutionary connection between them. Stars found below the canonical EHB at somewhat higher surface gravities show large RV-variations and a high RV-variability fraction, which is consistent with most of them being low-mass EHB stars or progenitors of low-mass helium white dwarfs in close binaries.} 


\maketitle

\section{Introduction \label{sec:intro}}

Hot subdwarf stars (sdO/Bs) constitute a prominent population of faint blue stars at high Galactic latitudes (Heber \cite{heber09,heber16}). With masses around $0.5\,M_{\rm \odot}$ and radii between $0.1\,R_{\rm \odot}$ and $0.3\,R_{\rm \odot}$ they are much smaller and of lower mass than hot main sequence stars of similar spectral types. Hot subdwarfs can form after main sequence stars such as the Sun expand and become red giants. This expansion stops as soon as helium burning starts in the red giant cores. Most of the observed subluminous B stars (sdBs) have been identified as extreme horizontal branch (EHB) stars burning helium in their cores (Heber \cite{heber86}). Although sdBs and sdOs occupy neighboring regions in the Hertzsprung-Russell diagram, they are quite different with respect to their chemical compositions. The atmospheres of sdBs are mostly helium poor and their helium abundances can be extremely low. Subluminous OB and O stars, on the other hand, show a large variety of helium abundances and can be divided in helium-poor sdOB/Os ($\log{n({\rm He})/n({\rm H})}\leq-1.0$), intermediate helium-rich iHe-sdOB/Os ($\log{n({\rm He})/n({\rm H})}=-1.0 ... 0.6$), and extreme helium-rich eHe-sdOB/Os ($\log{n({\rm He})/n({\rm H})}>0.6$).

\begin{figure*}[t!]
\begin{center}
	\resizebox{14cm}{!}{\includegraphics{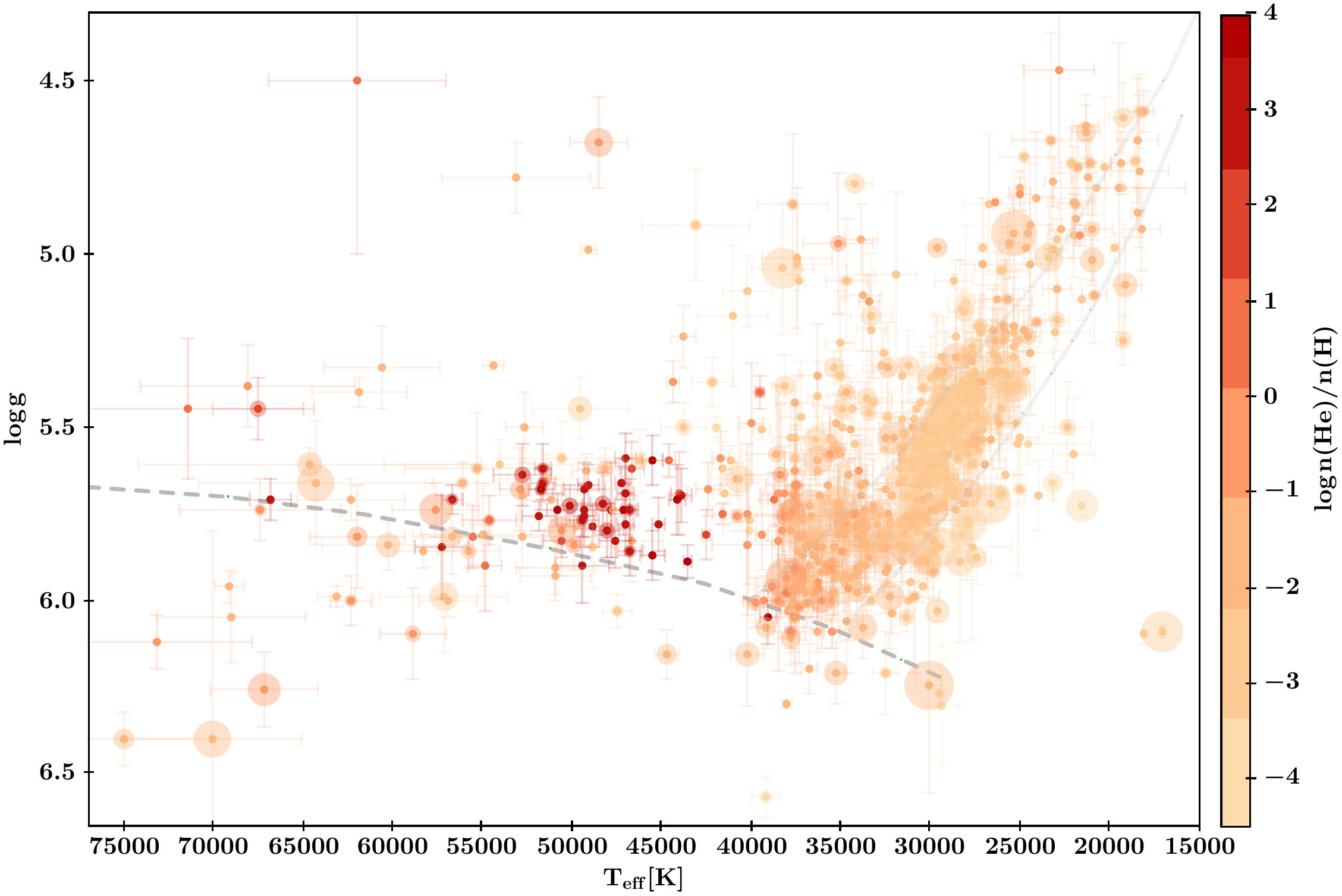}}
\end{center} 
\caption{$T_{\rm eff}-\log{g}$ diagram of the full sample of hot subluminous stars. The size of the symbols scales with $\Delta RV_{\rm max}$, the colour with the helium abundance from light orange to red. The EHB band (solid lines) is based on evolutionary tracks with subsolar metallicity ($\log{z}=-1.48$) from Dorman et al. (\cite{dorman93}). The helium main sequence (dashed line) is taken from Paczynski (\cite{paczynki71}).}
\label{tefflogg_all}
\end{figure*}

Hot subdwarfs can only be formed if the progenitor loses its envelope almost entirely after passing the red-giant branch (RGB) or if the ignition of He-burning occurs in an evolutionary stage late enough for the star to be already devoid of hydrogen. This is very difficult to explain in the context of single-star evolution, although single-star scenarios are still discussed. The attention shifted to binary evolution, when systematic surveys for radial velocity (RV) variable stars revealed a significant fraction (about one third) of the sdB stars to be members of close binaries (see Appendix~\ref{rvvar}). A similarly large fraction of the observed hot subdwarfs showed spectral features of cool main-sequence companions in wide binaries (see Stark \& Wade \cite{stark03} and references therein). 

Motivated by these discoveries, binary evolution scenarios were worked out (see Han et al. \cite{han02,han03} and references therein). Stable mass transfer to a main sequence companion via Roche lobe overflow (RLOF) has been proposed as major formation scenario for hot subdwarfs by Han et al. (\cite{han02,han03}) and composite sdB binaries with the predicted properties have later been discovered and studied (Vos et al. \cite{vos18} and references therein; Chen et al. \cite{chen13}; Vos et al. \cite{vos20}). The envelope stripping of intermediate and high-mass stars might lead to the formation of core helium-burning stars with higher masses as well (G\"otberg et al. \cite{goetberg18}).

To form close hot subdwarf binaries common envelope (CE) ejection is the only likely channel (see Han et al. \cite{han02,han03} and references therein). Several studies discovered and analysed hot subdwarfs in close binaries ($P\simeq0.03-30\,{\rm d}$) both with time-resolved spectroscopy and photometry (e.g. Copperwheat et al. \cite{copperwheat11}; Kawka et al. \cite{kawka15}; Kupfer et al. \cite{kupfer15}; Schaffenroth et al. \cite{schaffenroth19}). The majority of the mostly unseen companions are low-mass white dwarfs (WDs) and late main-sequence stars of spectral type M. A significant fraction of the sdBs are orbited by brown dwarfs (Schaffenroth et al. \cite{schaffenroth18} and references therein) and some have more massive compact companions (e.g. Geier et al. \cite{geier07,geier10}). Even three close double hot subdwarf binaries have been found (Sener \& Jeffery \cite{sener14}; Finch et al. \cite{finch19}; Reindl et al. \cite{reindl20}), some of which likely formed via double-core CE evolution (Justham et al. \cite{justham11}). Also the stripping by massive planets has been studied (Soker \cite{soker98}; Nelemans \& Tauris \cite{nelemans98}; Kramer et al. \cite{kramer20}).

\begin{figure*}[t!]
\begin{center}
	\resizebox{12cm}{!}{\includegraphics{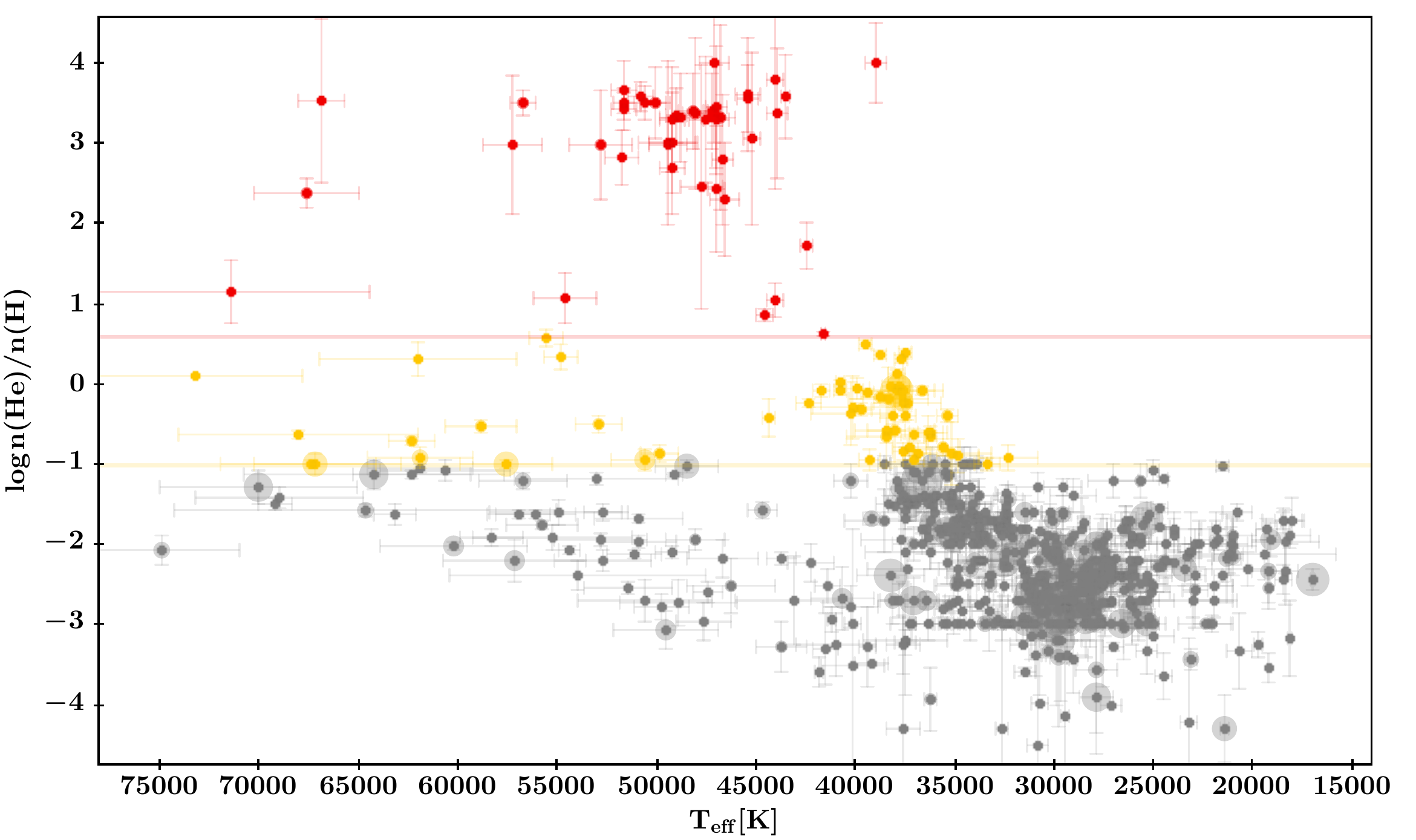}}
	\resizebox{12cm}{!}{\includegraphics{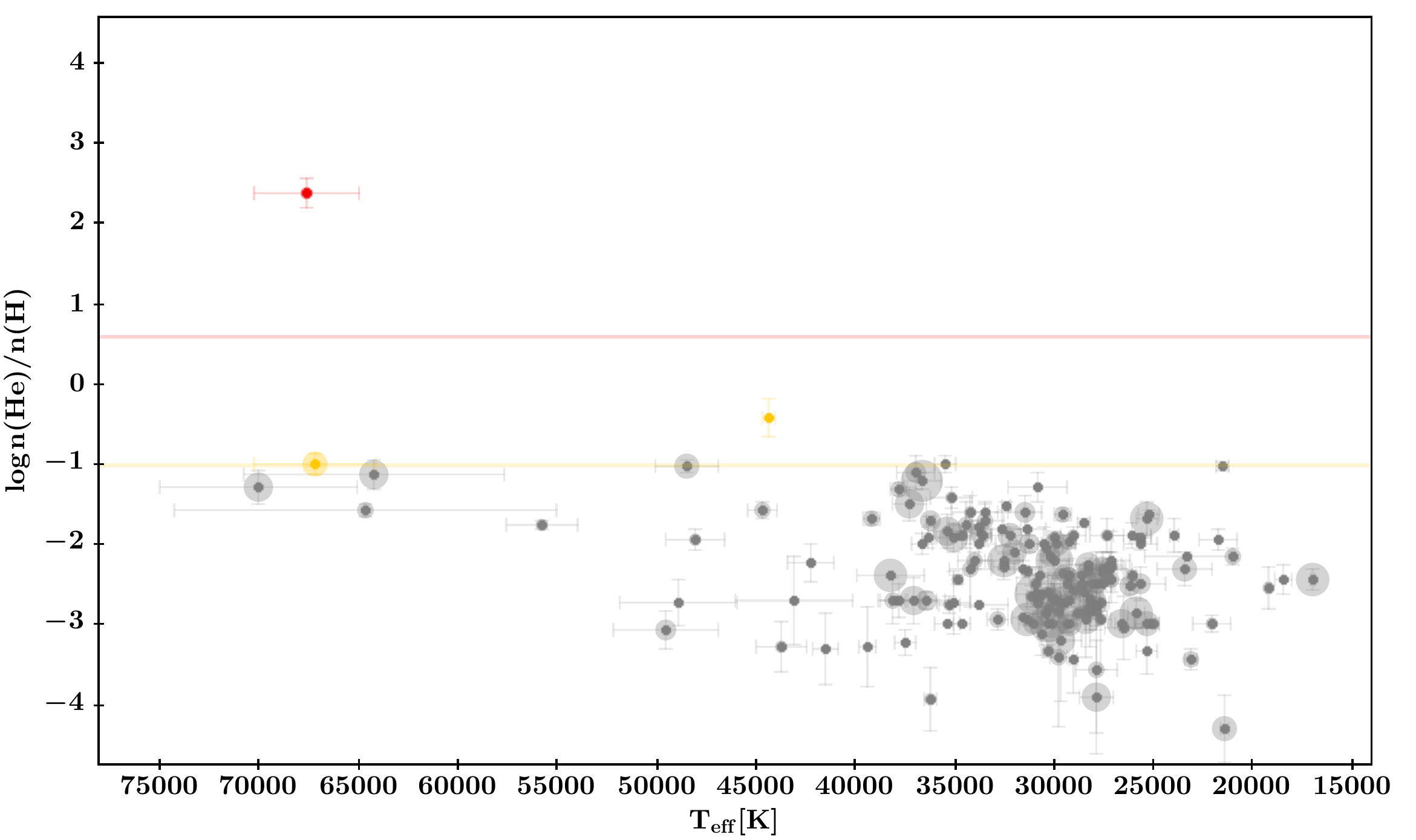}}
	\resizebox{11cm}{!}{\includegraphics{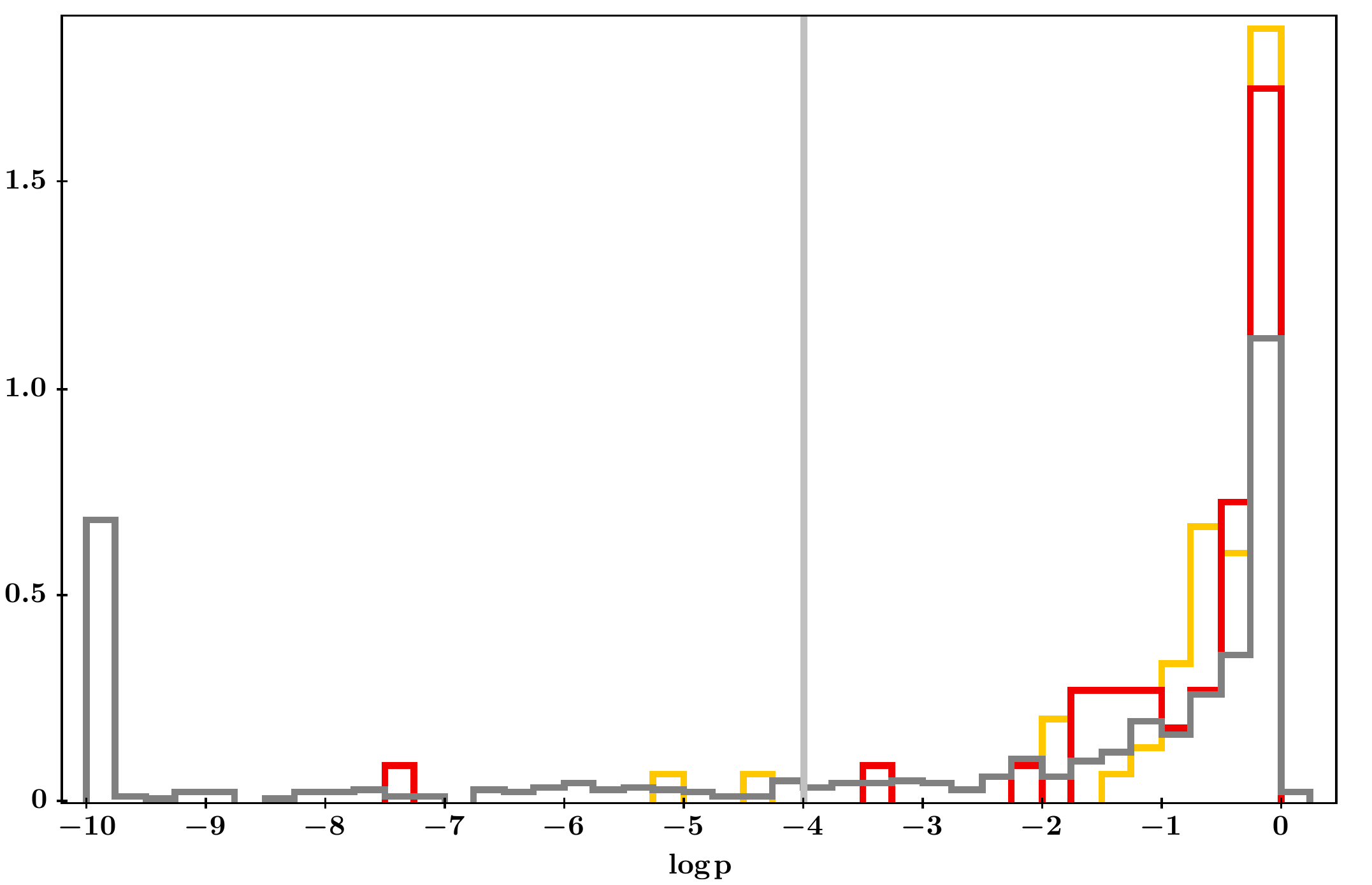}}
\end{center} 
\caption{Upper panel: $T_{\rm eff}-\log{n({\rm He})/n({\rm H})}$ diagram of the full sample. Helium-poor stars are marked in grey, intermediate helium-rich ones in orange and extreme helium-rich ones in red. The size of the symbols scales with $\Delta RV_{\rm max}$. Solar helium abundance is marked by the orange horizontal line, while the red line marks the transition between intermediate and extreme helium abundance. Middle panel: The same diagram for the stars showing significant RV variability ($\log{p}<-4.0$). Lower panel: $\log{p}$ distribution of the sample using the same colour-coding as in the upper panels. The significance level $\log{p}<-4.0$ is marked by a vertical line. The distribution has been limited to $\log{p}>-10$ and all objects with smaller $\log{p}$ have been stacked in the first bin for visualisation.}
\label{tefflogy_all}
\end{figure*}

About one third of the hot subdwarfs in the field show no indication for binarity and the fraction of those stars turns out to be much higher in globular clusters (GCs, Latour et al. \cite{latour18}). The merger of two helium white dwarfs (Webbink \cite{webbink84}) was proposed to explain the existence of those objects in the framework of binary evolution as well as mergers of a low-mass star or brown dwarf with a red-giant core  (Soker \cite{soker98}; Politano et al. \cite{politano08}; Kramer et al. \cite{kramer20}). Politano et al. (\cite{politano08}) predicted that this would lead to rapidly rotating single sdB stars. While the large majority of single sdBs are very slow rotators (Geier \& Heber \cite{geier12}), Geier et al. (\cite{geier11a,geier13b}) discovered two single sdBs, which are indeed fast rotators. Clausen et al. (\cite{clausen11}), however, propose that the coalescence of a helium white dwarf with a low-mass, hydrogen-burning star would create a star with a helium core and a thick hydrogen envelope that evolves into an sdB star after a few Gyrs, which would also naturally explain the sdBs' slow rotation rates.

Mixing processes during the merger of He-WDs are more consistent with the helium-rich composition of the He-sdOs. The location of the He-sdOs slightly blueward of the EHB (Str\"oer et al. \cite{stroeer07}; Nemeth et al. \cite{nemeth12}) matches with theoretical He-WD merger tracks (Zhang \& Jeffery \cite{zhang12}). In addition, the population of He-sdOs seems to consist mostly of single stars (see Appendix~\ref{rvvar}). Justham et al. (\cite{justham11}) proposed a merger channel involving a core-helium burning sdB and a He-WD for the formation of He-sdOs. 

Most recently, more evidence for the diverse types of merger formation channels has been found. Vos et al. (\cite{vos21}) discovered a hydrogen-rich sdB star surrounded by a gas disc indicative of a young merger product, Dorsch et al. (\cite{dorsch22}) presented a magnetic He-sdO perfectly consistent with the predictions of the He-WD merger channel, and finally Werner et al. (\cite{werner22}) reported the discovery of a whole new class of helium-rich sdOs showing extreme enrichments in carbon and oxygen, which are explained as the mergers of CO and He-WDs (Miller Bertolami et al. \cite{miller22}).  

Hot subdwarf stars might also form by mixing processes inside a single star without any binary interactions. Castellani \& Castellani (\cite{castellani93}) showed that sufficient mass loss on the RGB can lead to a delayed helium core flash (e.g. D'Cruz et al.  \cite{dcruz96}; Brown et al. \cite{brown01}). The flash drives convection and mixes the H-rich envelope into interior layers, where it is burned (Sweigart \cite{sweigart97a,sweigart97b}). The later the flash, the deeper the mixing leading to a higher helium abundance on the surface (Cassisi et al. \cite{cassisi03}; Lanz et al. \cite{lanz04}; Miller Bertolami et al. \cite{miller08}). An early hot flasher might explain single hydrogen-rich sdB stars, while late hot flashers might produce sdO/Bs enriched in helium (Naslim et al. \cite{naslim13}; Dorsch et al. \cite{dorsch19}). The deep mixing variant of a hot flasher scenario is most promising to explain the origin of the carbon-rich He-sdO stars (Heber et al. \cite{heber10}; Schindewolf et al. \cite{schindewolf18}). To explain EHB stars in globular clusters and elliptical galaxies, the enrichment of their parent population with helium has been proposed (e.g. Yi \cite{yi08} and references therein).

The general features visible in the $T_{\rm eff}-\log{g}$ plane seem to be consistent with the main evolutionary channels and evolutionary tracks (Dorman et al. \cite{dorman93}; Han et al. \cite{han02,han03}; Bloemen et al. \cite{bloemen14}; Xiong et al. \cite{xiong17}). Most sdBs are situated on the canonical EHB corresponding to a mass of $\sim0.47\,M_{\rm \odot}$, while the sdOB and sdO stars are located in the region of the post-EHB tracks, indicating an evolutionary link between those subclasses such that sdB stars will evolve to become sdOB and finally sdO stars before entering the WD cooling tracks. 

The He-sdO stars are concentrated in a region close to the helium main sequence where both the hot-flasher and merger tracks intersect. After the core helium-burning phase they evolve through a shell burning phase to hotter temperatures. Some of them might evolve to become very rare O(He) type stars and finally cool down as helium-rich WDs (Reindl et al. \cite{reindl14}). 

Only few sdBs are found to lie below the canonical EHB. All of them seem to be pre-He-WD objects in close binaries, which got stripped before the helium burning started in the core (e.g. Heber et al. \cite{heber03}). Those stars cross the EHB region while cooling down to become He-WDs and the evolutionary timescale for this process depends on their masses (Driebe et al. \cite{driebe98}; Althaus et al. \cite{althaus13}; Istrate et al. \cite{istrate16}).

Binary evolution scenarios also predict substructures to be present on the EHB (e.g. Han et al. \cite{han02,han03}; Xiong et al. \cite{xiong17}), due to the different formation channels. Hot subdwarfs from the CE-channel are predicted to occupy a region different from the hot subdwarfs formed via the RLOF- or merger-channels. However, those regions partly overlap and the observed samples are affected by selection effects (Han et al. \cite{han03}). In addition, Naslim et al. (\cite{naslim12}) proposed intermediate He-sdOBs with very peculiar metal-rich abundance patterns to be pre- instead of post-EHB objects, creating another degeneracy in the $T_{\rm eff}-\log{g}$ plane.

The degeneracies between the diverse evolutionary models make it difficult to link the various types of hot subdwarfs. Using other characteristic observable  properties seems to be a way forward. Usually the chemical composition is used for that purpose. However, the abundance patterns of hot subdwarfs are strongly affected by diffusion processes (O'Toole \& Heber \cite{otoole06}; Hu et al. \cite{hu11}; Michaud et al. \cite{michaud11}; Geier \cite{geier13}; Schneider et al. \cite{schneider18}; Byrne et al. \cite{byrne18}). Miller Bertolami et al. (\cite{miller08}) argue that He-sdO stars might even turn into sdBs due to gravitational settling in their atmospheres (see also Nemeth et al. \cite{nemeth12}; Luo et al. \cite{luo16}). For the hottest stages of post-EHB evolution also fractionated stellar winds are predicted to change the abundances (Unglaub \cite{unglaub08}; Krticka et al. \cite{krticka16}). 

\begin{figure*}[t!]
\begin{center}
	\resizebox{9cm}{!}{\includegraphics{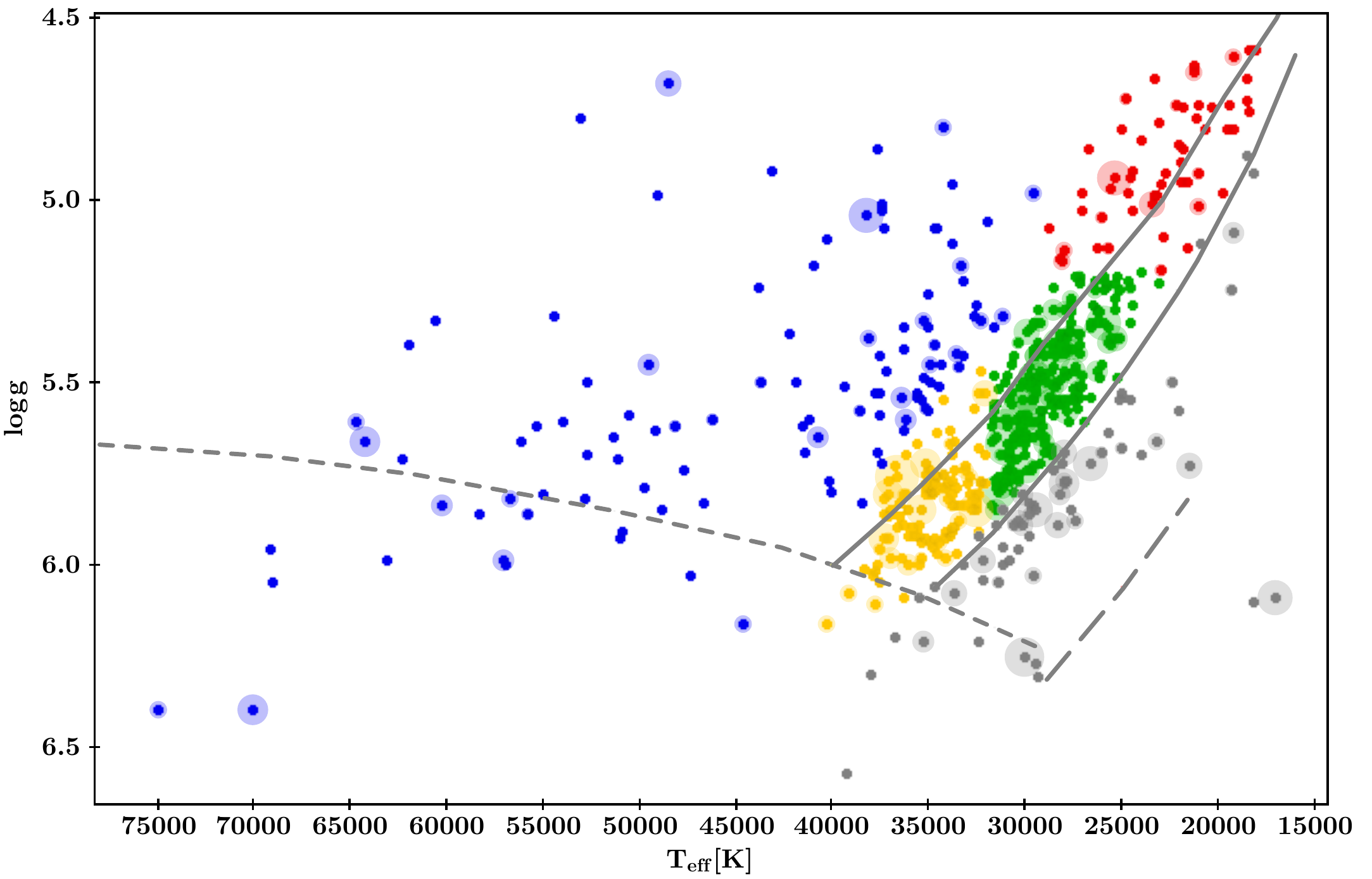}}
	\resizebox{9cm}{!}{\includegraphics{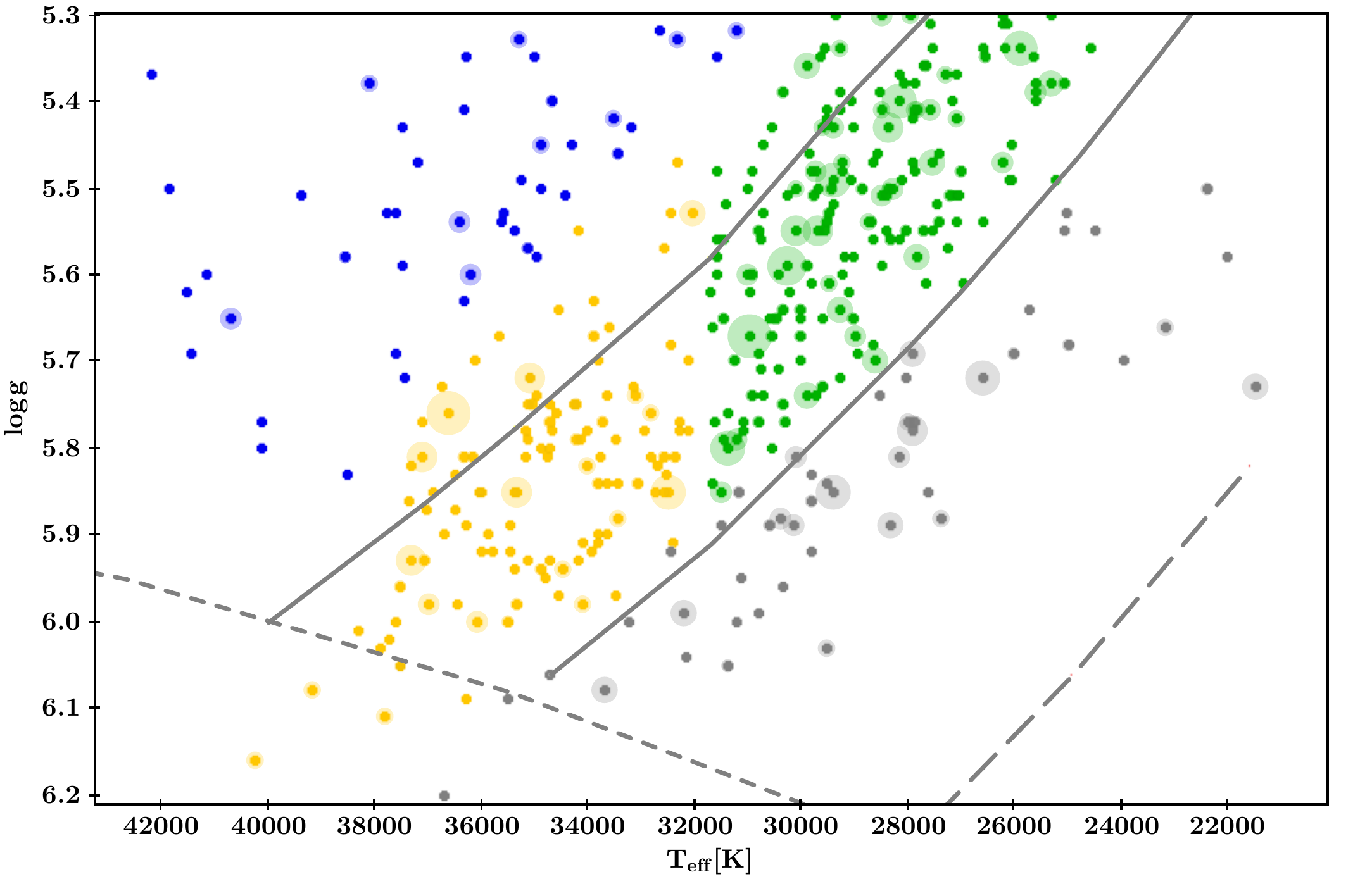}}
\end{center} 
\caption{Left panel: $T_{\rm eff}-\log{g}$ diagram similar to Fig.~\ref{tefflogg_all} of the sample of He-poor sdO/Bs with different regions marked by colour (EHB1 red, EHB2 green, EHB3 yellow, postEHB blue, bEHB grey) and the size of the symbol encoding $\Delta\,RV_{\rm max}$. In addition to the canonical EHB for a mass of $0.47\,M_{\rm \odot}$ (Dorman et al. \cite{dorman93}) the EHB for a low mass of $0.35\,M_{\rm \odot}$ (Han et al. \cite{han02}) is plotted as long-dashed grey line. Right panel: Close-up of the same diagram showing the division between the regions EHB2 and EHB3 in more detail.}
\label{tefflogg_selection}
\end{figure*}

\begin{figure*}[t!]
\begin{center}
    \resizebox{16cm}{!}{\includegraphics{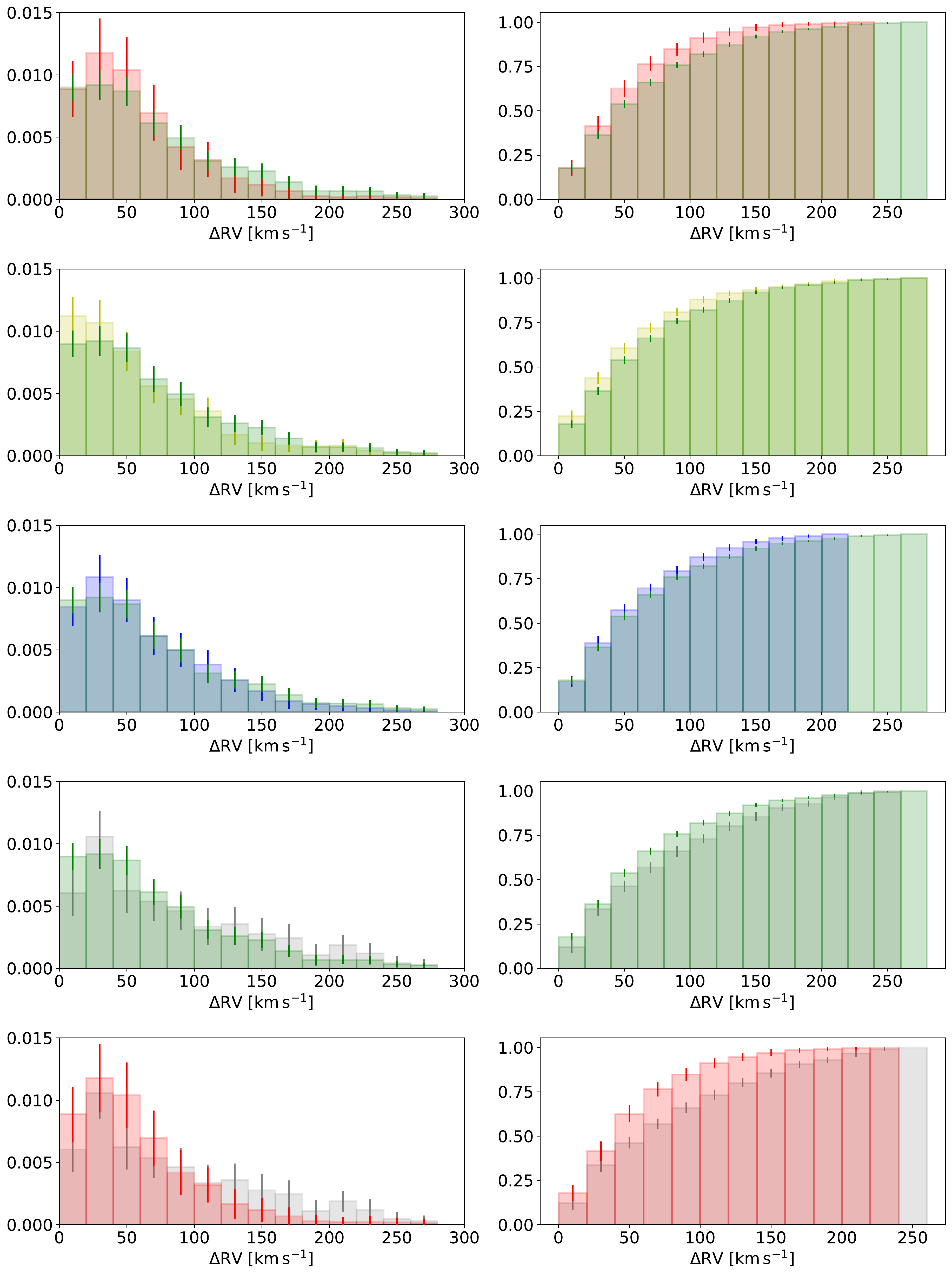}}
\end{center} 
\caption{Comparison of the normalised $\Delta\,RV_{\rm max}$ distributions (left panels) and the normalised cumulative $\Delta\,RV_{\rm max}$ distributions (right panels) in the different regions. From top to bottom: EHB1 (red) and EHB2 (green), EHB3 (yellow) and EHB2 (green), postEHB (blue) and EHB2 (green), bEHB (grey) and EHB2 (green), and bEHB (grey) and EHB1 (red). It has to be pointed out that the contribution of stars with high RV shifts tends to be visually overrepresented in the cumulative distributions and is better visible in the normalised distributions.}
\label{dRV_hist}
\end{figure*}

Since binary interactions play an important role for the formation of hot subdwarfs, we find many sdO/Bs in binary systems. In contrast to the chemical composition of the atmosphere, the properties of those binary systems are almost unaffected by the evolution in the sdO/B stage. Only the closest known sdB binaries will shrink significantly due to the emission of gravitational waves while evolving on and off the EHB and might even undergo another phase of mass transfer (see Kupfer et al. \cite{kupfer20} and references therein). For all other hot subdwarf binaries, the orbital parameters and the properties of the companions remain essentially unchanged. 

Most previous studies of the RV variability of hot subdwarf stars (Maxted et al. \cite{maxted01}; Napiwotzki et al. \cite{napiwotzki04a}; Morales-Rueda et al. \cite{morales03}; Copperwheat et al. \cite{copperwheat11}; Kawka et al. \cite{kawka15}) found that He-sdO/Bs and sdBs with composite spectra usually do not show significant RV variability (indications for significant RV variations of He-sdOs have been found by Green et al. \cite{green08} and Geier et al. \cite{geier15b,geier17a}), while the variability fractions of stars in the EHB and post-EHB region were rather inconsistent ranging from $30-70\%$. Selection effects such as preferential selection of the closest binaries or a biased mix of objects from different Galactic populations were proposed as possible reasons for these inconsistencies (see Appendix~\ref{rvvar} for details).

Recently, Pelisoli et al. (\cite{pelisoli20}) compared the fraction of wide, non-interacting hot subdwarf binaries with the respective fraction of the progenitor population of low-mass main sequence stars. The lack of such systems among the hot subdwarfs allowed the authors to provide observational evidence that pure single star evolution is very unlikely to result in the formation of sdO/B stars and that binary interactions are likely always required. 

Here we aim at using the close binary properties of hot subdwarf stars, which affect their RV variability, to put constraints on their formation and study their evolutionary links.

\section{A sample of hot subdwarfs with multi-epoch radial velocities} 

\subsection{Sample selection}

The sample studied here was compiled from the catalogue of spectroscopically classified hot subdwarfs (Geier et al. \cite{geier17a}) Data Release 2 (Geier \cite{geier20}).\footnote{The catalogue contains most likely AGB manqu\'e sdO/B type stars. The hotter sdO stars associated with post-AGB evolution have been excluded. For an early RV variability study of such objects see Reindl et al. (\cite{reindl16}).} The catalogue was crossmatched with the catalogues of the Sloan Digital Sky Survey Data Release 12 (SDSS DR12, Alam et al. \cite{alam15}) and the Large Sky Area Multi-Object Fibre Spectroscopic Telescope Data Release 5 (LAMOST DR5, Luo et al. \cite{luo19}) and stars with spectra taken at two or more epochs separated by a at least one day were selected. 

The study is restricted to sdO/Bs, which are not in composite binaries with companions visible in the spectrum. The main reason is that the atmospheric parameters of those double-lined binaries are more complicated to be determined and they are therefore often removed from analyses of larger samples (e.g. Luo et al. \cite{luo21}). Only few composite sdB binaries have been analysed in detail so far (e.g. Dorsch et al. \cite{dorsch21}; Nemeth et al. \cite{nemeth12,nemeth21}). It can also happen that the contribution of the companion is rather weak and not easily apparent when looking at the spectral features alone. The contribution to the continuum flux, however, can introduce significant systematic shifts in the atmospheric parameters usually towards higher temperatures and lower surface gravities. It is therefore important to exclude such a visible companion before performing a spectral analysis.

Known composite systems consisting of sdO/B stars and main-sequence F/G/K companions have therefore been excluded. In a second step we examined the spectral energy distributions (SEDs) of all the stars as described in Heber et al. (\cite{heber18}) to search for yet undetected composite binary systems and excluded them from our sample. This latter method is more sensitive than the inspection of optical spectra and allows to detect cool main sequence companions down late K-type. Our final sample is therefore restricted to single-lined stars. The companions in the detected close binary systems are either M-type main sequence stars, substellar objects, or compact objects such as white dwarfs.  

Although the selection functions of the sub-surveys conducted by SDSS and LAMOST are quite complicated, there should be no bias in favour or against RV-variable sdO/B stars within the observed sdO/B samples, because single-lined RV-variable and RV-constant sdO/Bs are indistinguishable in terms of colour or luminosity. The observing epochs can be considered as randomly distributed and the timespans between single measurements range from one day to several hundred days. Also in this respect the sample is completely unbiased against short or long periods of variability.

Some bias might be introduced by the different exposure times of the LAMOST and SDSS spectra. While the former spectra are exposed for $50\,{\rm min}$, the latter ones are exposed for $15\,{\rm min}$. For very short period binaries with high RV-variations the LAMOST spectra will be affected by orbital smearing, which should in general lead to an underestimation of the RV shifts. Biases are also introduced by the different sampling of the RV curves and the different number of epochs per object as well as the limited accuracy of our RV measurements (see Sect.~\ref{var}). All those biases, however, should affect the sub-samples we want to study similarly and therefore allow us to compare them in a meaningful way.

\subsection{Atmospheric parameters}

The atmospheric parameters effective temperature $T_{\rm eff}$, surface gravity $\log{g}$, and helium abundance $\log{n({\rm He})/n({\rm H)}}$ for the hydrogen-rich sdB and sdOB stars ($\log{n({\rm He})/n({\rm H})}<-1.0$) have been taken from the literature (Heber et al. \cite{heber87}; Saffer et al. \cite{saffer94}; Maxted et al. \cite{maxted01}; Edelmann et al. \cite{edelmann03}; Lisker et al. \cite{lisker05}; Str\"oer et al. \cite{stroeer07}; Charpinet et al. \cite{charpinet08}; Hirsch \cite{hirsch09}; \O stensen et al. \cite{oestensen10a,oestensen10b}; Nemeth et al. \cite{nemeth12}; Geier et al. \cite{geier13a,geier14,geier15b,geier17b}; Luo et al. \cite{luo16,luo19,luo21}; Lei et al. \cite{lei18,lei19,lei20}; Kepler et al. \cite{kepler19}; Hogg et al. \cite{hogg20}), if available. 

For the hydrogen-rich sdB and sdOB stars without any parameter determination in the literature, we fitted model spectra to the hydrogen and helium lines of the SDSS, BOSS or LAMOST spectra downloaded from the respective data archives using the SPAS routine (Hirsch \cite{hirsch09}) as described in Geier et al. (\cite{geier11b}). To increase the S/N, multiple spectra of one star have been shifted to rest wavelength and coadded. 

The quantitative spectral analysis was based on a new grid of model atmospheres and synthetic hydrogen and helium spectra calculated with the {\sc Atlas12} (Kurucz  \cite{kurucz96}), {\sc Detail}, and {\sc Surface} codes (Giddings \cite{giddings81}; Butler \& Giddings \cite{butler85}) that allowed us to treat some non-local thermodynamical equlibrium (NLTE) effects. All three codes were updated (Pryzbilla et al. \cite{przybilla11}; Irrgang et al. \cite{irrgang18}) and a grid covering the parameter range of hot subdwarfs was calculated (e.g. Schaffenroth et al. \cite{schaffenroth21}).

Owing to more pronounced NLTE-effects present in He-sdO/B stars, we re-determined the atmospheric parameters of all helium-enriched ($\log{n({\rm He})/n({\rm H})}\geq-1.0$) subdwarfs in our sample as described above using the SPAS routine together with a NLTE grid (Dorsch et al. \cite{dorsch19}). For this we computed non-LTE model atmospheres including hydrogen, helium, carbon, nitrogen, and silicon using the {\sc TLUSTY} and {\sc SYNSPEC} codes developed by Hubeny (\cite{hubeny88}) and Lanz \& Hubeny (\cite{lanz03}). 

Finally, sdO stars with temperatures higher than $70\,000\,{\rm K}$ have all been re-fitted with a grid of metal-free NLTE-models (Reindl et al. \cite{reindl16}). For the model calculations we used the T\"ubingen non-LTE model-atmosphere package (TMAP, Werner et al. \cite{werner03}; Rauch \& Deetjen \cite{rauch03}; Werner et al. \cite{werner12}). 

Our final sample consists of 646 stars with atmospheric parameter determinations and at least two epochs of spectroscopy.

\subsection{Radial velocities and criterion for variability}\label{var}

The radial velocities of the SDSS\footnote{https://www.sdss.org/dr12/algorithms/redshifts/} and LAMOST spectra provided in the archives are measured by matching template spectra to the data. In addition to the statistical uncertainties provided in the databases, systematic uncertainties have been determined to be about $5\,{\rm km\,s^{-1}}$ for all three spectrographs (Yanny et al. \cite{yanny09}; Bolton et al. \cite{bolton12}; Gao et al. \cite{gao15}). The LAMOST spectra have exposure times of $50\,{\rm min}$ and we downloaded the RVs for the hydrogen-rich sdO/B stars in our sample from the LAMOST DR5 using the Vizier database. 

The SDSS spectra taken with the SDSS and the upgraded BOSS spectrograph are co-added from at least three individual integrations with typical exposure times of $15\,{\rm min}$. Although in most cases those individual spectra are taken consecutively, sometimes also exposures taken with the same plate at different nights are stacked together to produce a co-added spectrum (Stoughton et al. \cite{stoughton02}). While this is not an issue for stars with small or no RV-variability (such as the majority of the main-sequence stars) or extragalactic objects, it can lead to erroneous RV measurements for objects showing high intrinsic RV variability such as close hot subdwarf binaries. 

We therefore did not use the RVs from the co-added spectra provided in the SDSS archive. Furthermore, for the helium-rich sdO/B stars in our sample, we could also not rely on the archival RVs, because the template databases do not contain proper He-rich templates. This can lead to a systematic shift in RV especially for hot stars, because the He\,{\sc ii} lines of the Pickering series are erroneously fitted as hydrogen Balmer lines. The most prominent example for this effect is the hypervelocity He-sdO US\,708, where the RV measurement from SDSS is off by more than $100\,{\rm km\,s^{-1}}$ (Geier et al. \cite{geier15a}). 

We downloaded all the individual SDSS and BOSS spectra as well as the LAMOST spectra of the He-sdO/Bs in the sample and measured the RVs using fixed sets of prominent spectral lines. For the hydrogen-rich sdO/Bs we used the hydrogen Balmer lines H\,$_{\beta}$, H\,$_{\gamma}$ and H\,$_{\delta}$. For the He-sdO/Bs we used the helium lines He\,{\sc i}\,4472, He\,{\sc i}\,4922, He\,{\sc ii}\,4541, He\,{\sc ii}\,4686, and He\,{\sc ii}\,5412. The lines were fitted with model spectra (Str\"oer et al. \cite{stroeer07}) by means of chi-squared minimization using the FITSB2 routine (Napiwotzki et al. \cite{napiwotzki04b}), and statistical $1\sigma$ errors were calculated. A systematic uncertainty of $5\,{\rm km\,s^{-1}}$ has been added in quadrature to our own measurements and the RVs provided in the LAMOST DR5 to obtain the final uncertainty of each measurement. Highly uncertain RV measurements with errors of more than $50\,{\rm km\,s^{-1}}$ have been discarded. In total, we measured and compiled 4311 single RVs for the 646 objects in our sample. 

The maximum RV variation $\Delta RV_{\rm max}$ was calculated as the difference between the maximum and the minimum RV of the star and the associated uncertainty was propagated from the uncertainties of the respective extreme measurements. To estimate the fraction of false detections produced by random outliers and calculate the significance of the measured RV variations we applied the method outlined in Maxted et al. (\cite{maxted01}), which was also used in similar studies by Geier et al. (\cite{geier15b,geier17a}), Latour et al. (\cite{latour18}) and Napiwotzki et al. (\cite{napiwotzki20}). 

For each star we calculate the inverse-variance weighted mean velocity from all measured epochs. Assuming this mean velocity to be constant, we calculate the $\chi^{2}$. Comparing this value with the $\chi^{2}$-distribution for the appropriate number of degrees of freedom we calculate the probability $p$ of obtaining the observed value of $\chi^{2}$ or higher from random fluctuations around a constant value. The detection of RV variability is considered to be significant, if the false-detection probability $p$ is smaller than $0.01\%$ ($\log{p}<-4.0$). Stars with false-detection probabilities ranging between $0.01\%$ and $5\%$ ($\log{p}=-4.0$ to $\log{p}=-1.3$) should be regarded as candidates, where follow-up spectroscopy might reveal significant RV variations in the future. 

The variability fractions in this study have been determined based on the number of objects with false detection probabilites smaller than $0.01\%$ ($\log{p}<-4.0$). Given the low-number statistics, the uncertainties of the variability fractions were calculated assuming a binomial distribution and indicate the $68\%$ confidence-level interval (see e.g. Burgasser et al. \cite{burgasser03}). It has to be pointed out, that the variability fractions have to be regarded as lower limits only, because the average uncertainty of our RV measurements is $18\,{\rm km\,s^{-1}}$ and in many cases we cannot exclude variations smaller than that. This is also the reason why we only use them for a differential analysis of the diverse subsamples studied here.

In Table~\ref{data} we provide the atmospheric parameters with literature reference, the number of RV epochs, weighted mean RVs, $\Delta RV_{\rm max}$, and false alarm probabilites $\log{p}$ for all the stars in our sample. The individual RVs will be published in a separate catalogue paper. The sample contains 164 stars with significant RV-variations. Only 19 of them are known binary systems with solved orbits. This sample increases the total number of known RV variable sdO/Bs by about a third.

\section{Radial velocity variability of hot subdwarfs}

Due to its large size, the full sample allows us to study the RV-variability of hot subdwarfs dependent on their location in the $T_{\rm eff}-\log{g}$- and $T_{\rm eff}-\log{n({\rm He})/n({\rm H})}$ diagrams and therefore dependent on their subtypes and evolutionary stages. Fig.~\ref{tefflogg_all} shows the $T_{\rm eff}-\log{g}$ diagram of the full sample, where the size of the symbols scales linearly with $\Delta RV_{\rm max}$ ranging from zero up to about $300\,{\rm km\,s^{-1}}$, while the colour scales with the helium abundance. It can be clearly seen that the distribution of $\Delta RV_{\rm max}$ is far from being isotropic and that rather complicated patterns exist. In the following we try to disentangle those substructures and their dependencies on the atmospheric parameters of the stars. An overview of our results is provided in Table~\ref{rvtable}.

\begin{table*}
\caption{\label{rvtable} RV variability of the sample and the comparison sample (M$\&$C) of Maxted et al. (\cite{maxted01}) and Copperwheat et al. (\cite{copperwheat11}).}
\begin{center}
\begin{tabular}{lllllll}
\hline
\noalign{\smallskip}
Subsample                  &   total   &  variable    & fraction [$\%$] &   total (M$\&$C) &  variable (M$\&$C) & fraction (M$\&$C) [$\%$] \\
\noalign{\smallskip}
\hline
\noalign{\smallskip}
He-poor                    &   539     &  161         & $30_{-2}^{+2}$  &   105            &  50                & $48_{-5}^{+5}$           \\
iHe-rich                   &   60      &  2           & $3_{-1}^{+4}$   &   $-$            &  $-$               & $-$                      \\
eHe-rich                   &   44      &  1           & $2_{-1}^{+5}$   &   $-$            &  $-$               & $-$                      \\
\noalign{\smallskip}
\hline
\noalign{\smallskip}
EHB1                       &   54      &  11          & $20_{-4}^{+6}$  &   4              &  1                 & $25_{-25}^{+10}$         \\
EHB2                       &   201     &  68          & $34_{-3}^{+3}$  &   48             &  29                & $60_{-7}^{+7}$           \\
EHB3                       &   116     &  26          & $22_{-3}^{+4}$  &   17             &  5                 & $29_{-13}^{+8}$          \\
postEHB                    &   107     &  28          & $26_{-4}^{+5}$  &   18             &  8                 & $44_{-12}^{+11}$         \\
bEHB                       &   61      &  28          & $46_{-6}^{+6}$  &   18             &  7                 & $39_{-12}^{+10}$         \\
\noalign{\smallskip}
\hline\hline
\end{tabular}
\end{center}
\end{table*}

\subsection{Helium abundance}

The upper panel of Fig.~\ref{tefflogy_all} shows the $T_{\rm eff}-\log{n({\rm He})/n({\rm H})}$ diagram of the full sample. The difference in RV-variability of hot subdwarfs with supersolar and subsolar helium abundance is striking. Many stars with low helium abundances (161 out of 539 objects) show moderate or high RV-variations, while only very few of the iHe- (two out of 60 objects) and eHe-sdO/Bs (one out of 44 objects) show any variation at all. The difference becomes even more striking when the sample is restricted to stars with significant RV variations (middle panel). The difference is also very pronounced when looking at the distribution of the $\log{p}$ values of the helium-poor and the helium-rich samples (Fig.~\ref{tefflogy_all}, lower panel). This is further confirmed when looking at the RV-variability fractions. While the He-poor sdO/Bs have a variability fraction of $30\pm2\%$, the variabilty fractions of intermediate He-rich ($3_{-1}^{+4}\%$) and extreme He-rich stars ($2_{-1}^{+5}\%$) are very small. 

Our results are in line with the preliminary results from the ESO Supernovae Ia Progenitor Survey (SPY) reported by Napiwotzki et al. (\cite{napiwotzki04a}) indicating a similarly low binary fraction for He-sdOs. The somewhat higher RV-variability fraction reported for the He-sdO/Bs from the Massive Unseen Companions to Hot Faint Underluminous Stars from SDSS (MUCHFUSS) survey might be related to the strongly biased selection procedure and the nature of the detected irregular variations (Geier et al. \cite{geier15b,geier17a}). The preliminary results obtained from the low-resolution sample of Green et al. (\cite{green08}) are also not consistent with the results presented here.

As can be seen in Fig.~\ref{tefflogg_all} the intermediate- and extreme He-sdO/Bs occupy neighbouring regions in the $T_{\rm eff}-\log{g}$-diagram with effective temperatures ranging from $\sim35\,000\,{\rm K}$ and $\sim55\,000\,{\rm K}$ and surface gravities ranging from $\log{g}\sim5.6$ to $6.0$. The clear difference in the RV variability with respect to the He-poor sdO/Bs excludes a direct evolutionary connection of those populations in general. He-sdO/Bs do not represent a later stage in the evolution of sdBs. This also means that diffusion processes are unlikely to form He-poor sdO/Bs as the progeny of more He-rich ones as proposed by Miller Bertolami et al. (\cite{miller08}) and Nemeth et al. (\cite{nemeth12}). Only confirmed single He-poor sdO/Bs might have formed in this way.

The similarities between iHe and eHe-sdO/Bs on the other hand allow for an evolutionary connection. Alternatively, both subtypes might just have formed in a similar way. Given that the most recent results of Pelisoli et al. (\cite{pelisoli20}) likely exclude pure single-star formation scenarios for sdO/Bs, the He-WD merger or the CE-merger channel might be a possibility. 

Also the fact that no eHe-sdO/B star in a close binary could be confirmed in this region of the $T_{\rm eff}-\log{g}$-diagram yet points in this direction. This  can be seen in Fig.~A\ref{tefflogg_lit}, where we show all known hot subdwarfs in close binary systems with known orbital and atmospheric parameters (Heber et al. \cite{heber03}; Silvotti et al. \cite{silvotti12}; Sener et al. \cite{sener14}; Kupfer et al. \cite{kupfer15} and references therein; Kawka et al. \cite{kawka15}; Schindewolf et al. \cite{schindewolf15}; Latour et al. \cite{latour14,latour16}; Hillwig et al. \cite{hillwig17}, priv. comm.; Kupfer et al. \cite{kupfer17a,kupfer17b,kupfer20,kupfer22}; Schaffenroth et al. \cite{schaffenroth18,schaffenroth21}; Vennes et al. \cite{vennes18}; Bell et el. \cite{bell19}; Ratzloff et al. \cite{ratzloff19,ratzloff20}; L\"obling \cite{loebling20}; Reindl et al. \cite{reindl20}; Pelisoli et al. \cite{pelisoli21}; Pawar et al. priv. comm.). 

Among the $\sim600$ known He-enriched hot subdwarfs there are only five solved close binaries. PG\,1544+488 is a double-lined binary consisting of two extreme He-sdBs with rather low temperatures of $32\,800\,{\rm K}$ and $26\,500\,{\rm K}$ (Sener \& Jeffery \cite{sener14}), Hen\,2$-$428 is a double-lined binary consisting of a iHe-sdOB and an sdOB with subsolar He-abundance (Reindl et al. \cite{reindl20}), and CPD$-$20\,1123 is a single-lined iHe-sdB ($T_{\rm eff}\sim25\,500\,{\rm K}$) with an unseen companion (Naslim et al. \cite{naslim12}; L\"obling \cite{loebling20}). The other two are the closest hot subdwarf binaries known:  OW\,J0741-2948 (Kupfer et al. \cite{kupfer17b}) and ZTF\,J2130+4420 (Kupfer et al. \cite{kupfer20}). The latter is currently transferring mass to its WD companion. 

The double He-sdO/Bs binaries can only be explained by specific and presumably quite rare formation channels (Justham et al. \cite{justham11}; Reindl et al. \cite{reindl20}). The iHe-sdOBs in the two very close binaries with WD companions are also peculiar objects, because of their likely quite massive progenitor stars. In this case core-helium burning is ignited under non-degenerate conditions and Kupfer et al. (\cite{kupfer20}) predicted such stars to be mildly enriched in helium. Due to their lower temperatures, the iHe-sdBs might not be related to the significantly hotter iHe- and eHe-sdOBs and sdOs.\footnote{The six He-sdO/Bs with significant RV variations discovered by Geier et al. (\cite{geier15b,geier17a}) could not be confirmed to be close binaries and their irregular variations likely have a different, yet unknown origin.} We therefore conclude that all He-rich sdO/Bs known in close binary systems likely belong to rare and peculiar sub-populations, whereas the large majority of the iHe- and eHe-sdO/Bs shows no convincing evidence for close binarity, which makes a merger origin very likely.

Fig.~\ref{tefflogy_all} (upper panel) also shows that further conclusions about the formation of hot subdwarfs based on their helium abundances are difficult. Luo et al. (\cite{luo19,luo20}) proposed that sdBs with $\log{n({\rm He})/n({\rm H})}<-2.2$ are formed via the stable RLOF channel and should therefore not show high RV-variability. In Fig.~\ref{tefflogy_all} (upper panel), however, many stars in this parameter range are clearly post-CE systems with high RV-variability.

\subsection{Position in the $T_{\rm eff}-\log{g}$-diagram}

Given that they likely represent a completely different population, we removed He-enriched sdO/Bs from the sample before studying the RV-variability properties  along the $T_{\rm eff}-\log{g}$-diagram. Guided by the structures seen in Fig.~\ref{tefflogg_all}, we have divided the $T_{\rm eff}-\log{g}$-diagram in five different regions shown in Fig.~\ref{tefflogg_selection}. The definition of the regions is based on visual inspection of the $T_{\rm eff}-\log{g}$-diagram. The three regions EHB1 (red), EHB2 (green), and EHB3 (yellow) are located on the EHB. Instead of relying on the uncertain location of the EHB band provided by evolutionary tracks, we took the density of the objects as a proxy to define the EHB assuming that the density is higher on than off the EHB. The region postEHB (blue) covers the stars located above the EHB. Finally, the region bEHB (grey) covers all the objects below the canonical EHB. 

\subsubsection{Extreme horizontal branch}

As can be seen in Fig.~\ref{tefflogg_all} and Fig.~\ref{tefflogg_selection} it turns out that the $\Delta RV_{\rm max}$ distribution along the EHB looks quite inhomogeneous.\footnote{Green at al. (\cite{green08}) presented a $T_{\rm eff}-\log{g}$-diagram (Fig.~3) where the symbols are scaled with standard deviation of RV measurements instead of $\Delta RV_{\rm max}$. The structures seen on the EHB are very similar to our results.} There seems to be a difference between the most populated region EHB2 (201 objects) and the sparsely populated region EHB1 (54 objects), which are divided at an effective temperature of $\sim24\,000\,{\rm K}$. Comparing the $\Delta\,RV_{\rm max}$-distributions  as well as the normalised cumulative $\Delta\,RV_{\rm max}$-distributions of both regions (see Fig.~\ref{dRV_hist} top panel) it can be seen that the RV shifts in region EHB1 are smaller. To assess the statistical significance of this comparison, we performed a Kolmogorov-Smirnow (KS) test. In this way also the uncertainties of the single measurements and the quite different sample sizes are taken into account. We determined a $p_{\rm KS}$-value of $0.12$, meaning that the hypothesis that both samples have the same distribution can be rejected with a probability of $88\%$. However, this value is still lower than the typical $95\%$ threshold, so the difference between both distributions is not significant. The variability fraction in region EHB2 is $34\pm3\%$ compared to only $20_{-4}^{+6}\%$ in region EHB1.

The distinction between the regions EHB2 and EHB3 (116 objects) is shown in Fig.~\ref{tefflogg_selection} (right panel). There seems to be a region devoid of stars around $T_{\rm eff}\sim33\,000\,{\rm K}$ and $\log{g}\sim5.7$. As can be seen in Fig.~\ref{dRV_hist} (second panels from the top) there is a less pronounced difference in the $\Delta RV_{\rm max}$ distributions of the regions EHB2 and EHB3. Again the KS-test ($p_{\rm KS}=0.18$) provides an indication for a difference between the two regions, which turned out to be not statistically significant. The RV-variability fraction in region EHB3 ($22_{-3}^{+4}\%$) is smaller than in region EHB2, which might be due to a contamination of non-RV-variable iHe-sdOBs with inaccurate helium abundance determinations, which are also located in this region. Alternatively, single iHe-sdOBs might evolve to become sdOBs due to diffusion processes in their atmospheres as proposed by Miller Bertolami et al. (\cite{miller08}).

The differences between regions EHB1 and EHB2 are more pronounced. The higher fraction of apparently single stars and the smaller RV-variations in region EHB1 compared to region EHB2 are indications for lower-mass companions and/or longer orbital periods. The binary population synthesis models of Han et al. (\cite{han03}) actually predict a desert of RV-variable systems for hot subdwarfs in region EHB1, because neither the first nor the second CE-channel in their simulations are able to form stars with hydrogen envelopes thick enough to correspond to such cool temperatures and low surface gravities (see Sect.~\ref{models}). According to their predictions, all those sdBs should come from the stable-RLOF channel and therefore show just small RV-variations undetectable by this study. However, in our pre-selection we discarded sdBs with cool companions detectable in their SEDs, which have been identified as the normal post-RLOF sdB systems. The binary properties in region EHB1 are therefore hard to explain by binary evolution theory and follow-up observations are needed to increase the statistical significance of these results.

\subsubsection{Above and beyond the extreme horizontal branch}

The region postEHB (107 objects) in Fig.~\ref{tefflogg_selection} is situated above and bluewards of the EHB. The exact location of the terminal age EHB (TAEHB) depends on the metallicity and the core mass of the stars and is therefore not very well defined. The distinction here has therefore to be regarded as a qualitative one only. Most stars in this regions should be evolved post-EHB stars with ongoing He-shell burning. 

Comparing the $\Delta RV_{\rm max}$ distribution we can now probe one of the most important evolutionary connections proposed for hot subdwarfs. Hot subdwarf B stars located on the EHB are proposed to evolve to post-EHB sdOB and sdO stars. In Fig.~\ref{dRV_hist} (middle panels) the $\Delta RV_{\rm max}$ distribution of region EHB2 is compared to the one of region postEHB. The distributions look similar. The KS-test with a high $p_{\rm KS}=0.47$ confirms the similarity of both distributions further and the RV variability fraction of region postEHB ($26_{-4}^{+5}\%$) is consistent with the one of region EHB2 within the uncertainties.\footnote{This is inconsistent with the reported mismatch between the RV-variability fractions of sdBs and sdOBs from the MUCHFUSS project, which was likely caused by selection effects (Geier et al. \cite{geier15b,geier17a})}

\subsubsection{Below the extreme horizontal branch}

The region bEHB (61 objects) is situated below the canonical EHB at somewhat higher surface gravities and covers the whole temperature range (see Fig.~\ref{tefflogg_selection}). It might be populated by sdO/Bs of different origins. Core helium-burning stars slightly below the canonical ZAEHB for low metallicity populations (such as the thick disk), which is indicated in Fig.~\ref{tefflogg_selection}, might belong to the more metal-rich thin-disk population (e.g. Dorman et al. \cite{dorman93}). 

Alternatively, those stars might have masses lower than the canonical mass. Intermediate-mass stars ($2-3\,M_{\rm \odot}$), which ignite core helium-burning non-degenerately, can evolve to become hot subdwarfs with masses down to $0.3\,M_{\rm \odot}$, which is the minimum mass for the ignition of He-burning. The EHB for such low-mass stars is also shifted towards higher surface gravities (Han et al. \cite{han02}; Bloemen et al. \cite{bloemen14}; Wu et al. \cite{wu18}, see Fig.~\ref{tefflogg_selection}). Due to the higher progenitor masses it is also expected that more massive companions in combination with a deeper spiral-in are necessary to eject the more tightly bound envelopes of the red giants. This scenario was proposed the explain the closest known sdB+WD binaries (e.g. Geier et al. \cite{geier13c}).

Finally, pre-He-WDs with no active He-burning are also found in this region. Since they are not connected to the EHB at all, they can be situated anywhere in the region bEHB. However, due to the quite different evolutionary timescales, low-mass pre-He-WDs ($\sim0.25\,M_{\rm \odot}$) at the low temperature end shown here should be more frequent than their hotter siblings with higher masses ($0.3-0.35\,M_{\rm \odot}$, Driebe et al. \cite{driebe98}; Althaus et al. \cite{althaus13}; Istrate et al. \cite{istrate16}). However, despite their shorter evolutionary times, the more massive pre-He-WDs might still be present, if more of them are formed in the first place. All pre-He-WDs are expected to be formed by binary mass-transfer and should have companions. The concentration of the hotter bEHB objects ($>25000\,{\rm K}$) below the EHB (see Fig.~\ref{tefflogg_selection}) might indicate that most of them are low-mass EHB objects, while the more widely distributed cooler objects might follow the pre-He-WD cooling tracks.

In Fig.~\ref{dRV_hist} (second panels from the bottom) the $\Delta RV_{\rm max}$ distribution of region bEHB is compared to the one of region EHB2. The distribution of the bEHB region is similar in terms of the high fraction of RV-variable objects, but somewhat flatter, indicating a higher fraction of binary systems with high RV-amplitudes. However, also in this case the KS-test ($p_{\rm KS}=0.33$) does not reveal a statistically significant difference between the two samples. In contrast to that, such a difference is found when comparing region EHB1 and region bEHB ($p_{\rm KS}=0.05$) and also the $\Delta RV_{\rm max}$ distributions (Fig.~\ref{dRV_hist} bottom panels) look different. 

The RV-variability fraction of $46\pm6\%$ in region bEHB is higher than the ones in regions EHB2 and EHB1, consistent with an intrinsically high close binary fraction. If confirmed, the somewhat higher RV-amplitudes would be consistent with a sample of binaries, where the visible primaries are less massive EHB stars or pre-He-WDs probably with higher mass companions and therefore show higher RV-variations. Also shorter orbitals periods are possible, because the CE-ejection must have happened before the progenitor reached the tip of the RGB to form a pre-He-WD and the higher binding energy of the envelope favors a deeper inspiral of the companion.

\begin{figure*}[t!]
\begin{center}
	\resizebox{14cm}{!}{\includegraphics{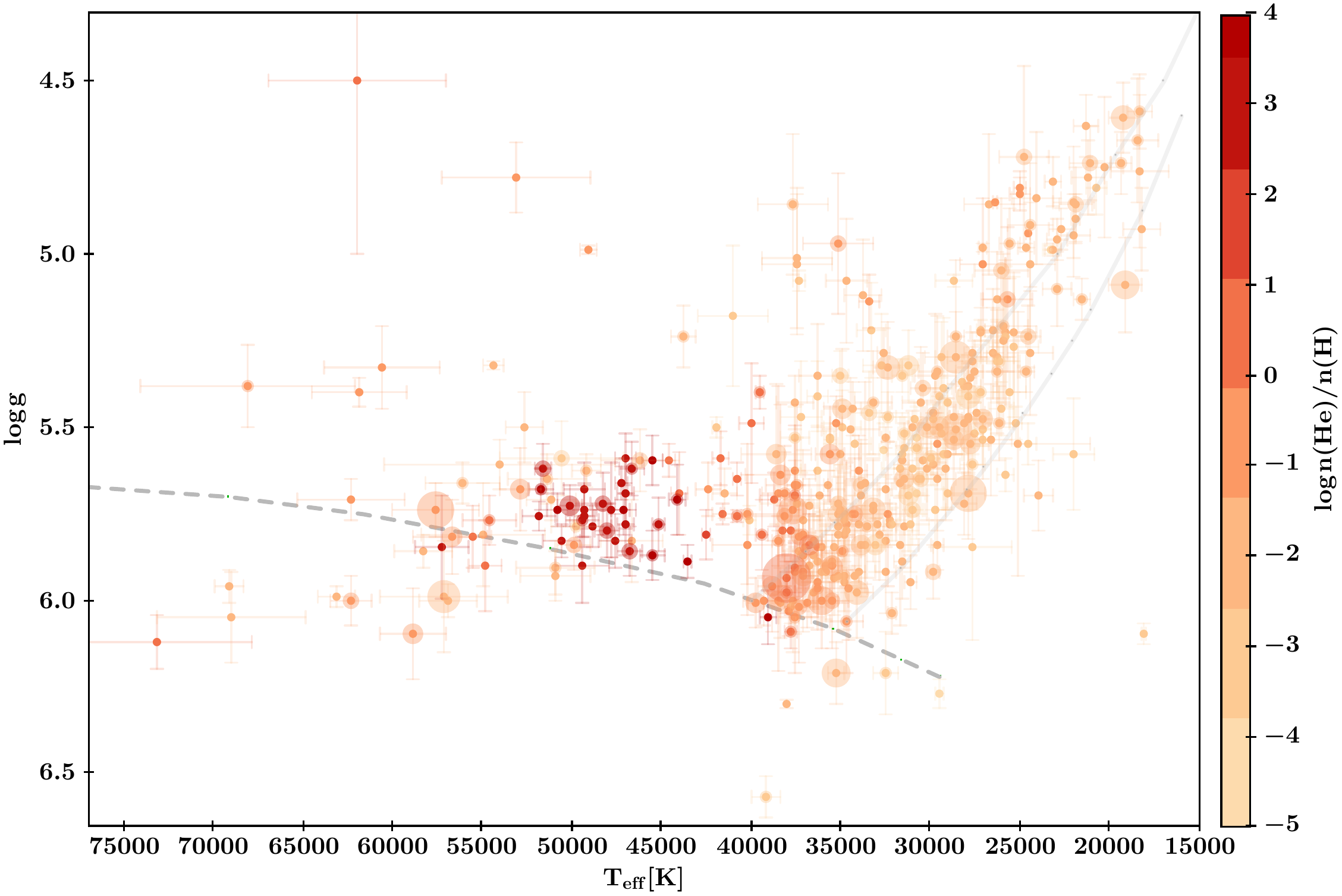}}
\end{center} 
\caption{$T_{\rm eff}-\log{g}$ diagram similar to Fig.~\ref{tefflogg_all} of the sample of hot subluminous stars showing no significant RV variations.}
\label{tefflogg_const}
\end{figure*}

\subsubsection{Comparison with the RV variability studies in the literature}

As can be seen in Appendix~\ref{rvvar}, the results from previous RV variability studies of sdO/Bs yielded quite inhomogeneous results and are due to different systematic biases difficult to be compared in a quantitative way. The sample most similar to the one studied here has initially been published by Maxted et al. (\cite{maxted01}) and later been extended and completed by Copperwheat et al. (\cite{copperwheat11}). Therefore we refer to it as the M$\&$C sample.\footnote{In an additional paper of the series by Morales-Rueda et al. (\cite{morales03}), binary solutions of many stars in the sample were published. However, also binaries discovered in other projects were included.} The authors used the same method to determine the significance of the RV variations as we did in this study, but their RV accuracy was better (down to $\sim2\,{\rm km\,s^{-1}}$). The main differences are the smaller sample size (about one fourth of our sample) and the lack of He-rich stars. Apart from that, the M$\&$C sample is well suited as a comparison sample to check whether the inhomogeneities we found for the He-poor sdO/Bs are also present at higher RV accuracy.

Initially, only a subset of the M$\&$C sample had atmospheric parameters published (Saffer et al. \cite{saffer94}; Maxted et al. \cite{maxted01}; Morales-Rueda et al. \cite{morales03}; Copperwheat et al. \cite{copperwheat11}). To extent this, we performed a literature research and found atmospheric parameters for two thirds (105 stars) of the M$\&$C sample (Heber et al. \cite{heber84}; Edelmann et al. \cite{edelmann04}; Lisker et al. \cite{lisker05}; \O stensen et al. \cite{oestensen10b}; Geier et al. \cite{geier11a,geier13a,geier17b}; Nemeth et al. \cite{nemeth12}; Luo et al. \cite{luo21}). 

Since a lot of the binaries discovered by M$\&$C were followed-up and the orbital parameters determined, this introduces a bias in the $\Delta RV_{\rm max}$ distribution with respect to our sample. This is why we refrained from comparing those. Instead we just determined the RV variability fractions of the M$\&$C sample (after removing the composite binaries) for the He-poor subsamples EHB1, EHB2, EHB3, postEHB and bEHB as defined above (see Table~\ref{rvtable}). Likely due to the higher RV accuracy, the overall RV variability fraction of the He-poor M$\&$C sample ($48\pm5\%$) is significantly higher than the one of our He-poor sample ($30\pm2\%$). This is even more pronounced in the region EHB2 (M$\&$C $60\pm7\%$ compared to $34\pm3\%$) and also marginally in region postEHB, while the variability fractions of the less populated regions EHB1, EHB3 and bEHB are consistent within the substantial uncertainties. However, there are clear indications that the binary fraction in region EHB2 of the M$\&$C sample is significantly higher than the one of region EHB3 confirming the trend seen in our sample. The comparison with region EHB1 is less obvious, because of the very small sample size, but also indicates a difference similar to the one in our sample.

\subsubsection{Comparison with binary population synthesis models}\label{models}

Han et al. (\cite{han02,han03}) performed an extensive binary population synthesis (BPS) study of hot subdwarf formation through binary interactions and compared the results to the observed samples of binary sdO/Bs available back then. Assuming that all hot subdwarfs are formed by binary evolution either via CE ejection, stable RLOF or He-WD mergers the distribution of the stars in the $T_{\rm eff}-\log{g}$ diagram depends on the input parameters of the BPS simulations such as the efficiency of the CE ejection. 

By comparison with the observed population, Han et al. (\cite{han03}) favoured a model population with solar metallicity ($Z=0.02$), a rather high CE ejection efficiency ($\alpha_{\rm CE}=0.75$) and an equally high thermal contribution to the binding energy of the envelope ($\alpha_{\rm th}=0.75$). Lisker et al. (\cite{lisker05}) compared the same simulations to a sample of hydrogen-rich sdBs and sdOBs from the SPY survey and achieved a better match with a model population of subsolar metallicity ($Z=0.004$) and with lower $\alpha_{\rm CE}=0.5$ as well as $\alpha_{\rm th}=0.5$.

Due to the inhomogeneous selection of our sample, we refrain from performing a quantitative comparison with the results of Han et al. (\cite{han03}) and leave that to future studies ideally based on updated BPS models. Instead we focus on obvious features and compare our $T_{\rm eff}-\log{g}$ diagram (Fig.~\ref{tefflogg_all})  with the models of Han et al. (\cite{han03}) as visualized in Fig.~17 of Lisker et al. (\cite{lisker05}). The model populations are corrected for observational selection effects in the sense that composite systems with MS companions of K- or earlier types have been removed, which matches our selection of single-lined stars very well. 

A prominent and distinctive feature of the models is the extend of the EHB towards low temperatures, which turns out to be very sensitive to the CE ejection parameters. The more efficient the CE ejection, the more objects are found at low temperatures. This corresponds to the low temperature edge of region EHB2 in our sample, while region EHB1 is not populated in the model populations at all. It can be clearly seen that our sample strongly favours models with very few stars cooler than $\sim25\,000\,{\rm K}$. This is consistent with the result from Lisker et al. (\cite{lisker05}) and indicates low values of $\alpha_{\rm CE}=0.5$, $\alpha_{\rm th}=0.5$ and $Z=0.004$.

\section{Summary and conclusions}

We performed a RV-variability analysis of a large sample of 646 hot subdwarfs with 4311 multi-epoch radial velocities from SDSS and LAMOST spectra. Atmospheric parameters and RVs were taken from the literature, if available and appropriate. For stars with archival spectra and no literature values, we determined the parameters by fitting model atmospheres. In particular, due to the higher  systematic uncertainties, we redetermined the atmospheric parameters and RVs for all the He-enriched sdO/Bs. This large sample allowed us to study RV-variability as a function of the location in the $T_{\rm eff}-\log{g}$- and $T_{\rm eff}-\log{n({\rm He})/n({\rm H})}$ diagrams in a statistically significant way. As diagnostics we used the fraction of RV-variable stars and the distributions of the maximum RV variations $\Delta RV_{\rm max}$. Both indicators turned out to be inhomogeneous across the studied parameters ranges. We discovered 145 new hot subdwarf stars with significant RV variations. The results of this study allowed us to draw several conclusions:

\begin{itemize}
\setlength\itemsep{1em}

\item {\bf Most He-rich sdO/Bs are single stars:} Both iHe- and eHe-sdO/Bs do not show significant RV-variability fractions in contrast to the He-poor sdO/Bs. They are therefore very likely single stars. This confirms preliminary results (Napiwotzki et al. \cite{napiwotzki04a}) for the first time a in statistically significant way. 

\item {\bf He-poor and He-rich sdO/Bs are not related:} The completely different behaviour of the He-poor and the He-rich hot subdwarfs led us to the conclusion that both subtypes are very likely not evolutionarily related. The iHe- and eHe-sdOB/Os on the other hand likely constitute one population. 

\item {\bf He-rich sdO/Bs are likely formed by mergers:} Since recent results from the study by Pelisoli et al. (\cite{pelisoli20}) indicate that single star evolution is unlikely for the formation of hot subdwarfs in general, we conclude that this population is formed via the merger channel. The most recent discovery of several hot subdwarf merger candidates is also in line with our conclusion (Vos et al. \cite{vos21}; Dorsch et al. \cite{dorsch22}; Werner et al. \cite{werner22}).

\item {\bf There are indications for inhomogeneous RV-variability of the He-poor sdO/Bs:} Most of the RV-variable post-CE binaries are found in a well defined region on the EHB, which has been predicted by binary population synthesis models. Hot subdwarfs with temperatures cooler than $\sim24\,000\,{\rm K}$ tend to show smaller RV-variations and a smaller RV-variability fraction. This small number of objects might constitute yet another subpopulation of binaries with longer periods and late-type or compact companions. The RV-variability properties of the EHB and corresponding post-EHB population of the He-poor hot subdwarfs match and confirm the predicted evolutionary connection between them. Stars found below the EHB show large RV-variations and a significant RV-variability fraction, which is consistent with them being either low-mass EHB stars or non core helium-burning objects and therefore progenitors of He-WDs. Although some of the apparent inhomogeneities of RV-variability in the $T_{\rm eff}-\log{g}$-diagram turned to be not statistically significant, we still think that they might be real. A comparison with the sample of M$\&$C points in this direction. Comparing the distribution of objects in the $T_{\rm eff}-\log{g}$-diagram with BPS models by Han et al. (\cite{han03}), we find indications for a rather low efficiency of CE-ejection. More sophisticated BPS models and a proper treatment of the selection biases are needed to make progress in this direction. 

\item {\bf Apparently single sdO/Bs are common:} A persistent riddle are the many sdO/B stars, which do not show any significant RV variability ($>50\%$ of the single-lined He-poor sdO/Bs and $\sim97\%$ of the He-rich sdO/Bs) and also no hints of a cool companion in their SEDs. As can be seen in Fig.~\ref{tefflogg_const}, those objects are found all over the $T_{\rm eff}-\log{g}$-diagram. Although a fraction of them might be explainable as binaries with small RV-variations due to long periods or low inclinations, it is very unlikely that all of them are binaries, since many sdO/Bs have been confirmed to be single in the literature based on high-resolution spectroscopy already (e.g. Silvotti et al. \cite{silvotti20}). For the iHe- and eHe-sdO/Bs the merger channel provides a convenient formation scenario, which also explains their location in the $T_{\rm eff}-\log{g}$-diagram. For the cooler He-poor sdO/Bs, other types of stellar mergers might be possible formation channels (Politano \cite{politano08}; Clausen \& Wade \cite{clausen11}; Hall \& Jeffery \cite{hall16}). Some or all of the single He-rich sdO/Bs might still turn into He-deficient stars due to diffusion processes as proposed by Miller Bertolami et al. (\cite{miller08}). Whether the even higher fraction of apparently single sdO/Bs in GCs (Latour et al. \cite{latour18}) could also be explained along those lines or might be caused by alternative formation channels (e.g. Yi \cite{yi08}) is still unclear and requires further studies.

\end{itemize}

Our results open many different routes for more detailed studies. Since the RV accuracy of the medium-resolution survey spectra used here is limited, follow-up studies with high-resolution spectra are needed to estimate the true fractions of non-variable hot subdwarfs and their distribution in the parameter space discussed here. More orbital solutions of binaries at the high and low temperature ends of the EHB as well as the regions below the EHB are needed to characterize those subpopulations. Furthermore, kinematic studies should be performed to determine the Galactic population properties of this sample. 

\begin{acknowledgements}

We would like to thank Thomas Kupfer, Veronika Schaffenroth and Evan Bauer for useful advice. We thank the anonymous referee for helpful comments and suggestions. SG acknowledges funding from the German Academic Exchange Service
(DAAD PPP USA 57444366) for this project and would like to thank the host institution Texas Tech University for the hospitality. IP was partially funded by the Deutsche Forschungsgemeinschaft under grant GE2506/12-1 and by the UK's Science and Technology Facilities Council (STFC), grant ST/T000406/1. UH and MD acknowledge funding by the Deutsche Forschungsgemeinschaft (DFG) through grants IR190/1-1, HE1356/70-1 and HE1356/71-1.

This research made use of TOPCAT, an interactive graphical viewer and editor for tabular data Taylor (\cite{taylor05}). This research made use of the SIMBAD database, operated at CDS, Strasbourg, France; the VizieR catalogue access tool, CDS, Strasbourg, France. 

This work has made use of data from the European Space Agency (ESA) mission {\it Gaia} (https://www.cosmos.esa.int/gaia), processed by the {\it Gaia} Data Processing and Analysis Consortium (DPAC, https://www.cosmos.esa.int/web/gaia/dpac/consortium). Funding for the DPAC has been provided by national institutions, in particular the institutions participating in the {\it Gaia} Multilateral Agreement.

Guoshoujing Telescope (the Large Sky Area Multi-Object Fiber Spectroscopic Telescope LAMOST) is a National Major Scientific Project built by the Chinese Academy of Sciences. Funding for the project has been provided by the National Development and Reform Commission. LAMOST is operated and managed by the National Astronomical Observatories, Chinese Academy of Sciences.

Funding for the SDSS and SDSS-II has been provided by the Alfred P. Sloan Foundation, the Participating Institutions, the National Science Foundation, the U.S. Department of Energy, the National Aeronautics and Space Administration, the Japanese Monbukagakusho, the Max Planck Society, and the Higher Education Funding Council for England. The SDSS Web Site is http://www.sdss.org/. The SDSS is managed by the Astrophysical Research Consortium for the Participating Institutions. The Participating Institutions are the American Museum of Natural History, Astrophysical Institute Potsdam, University of Basel, University of Cambridge, Case Western Reserve University, University of Chicago, Drexel University, Fermilab, the Institute for Advanced Study, the Japan Participation Group, Johns Hopkins University, the Joint Institute for Nuclear Astrophysics, the Kavli Institute for Particle Astrophysics and Cosmology, the Korean Scientist Group, the Chinese Academy of Sciences (LAMOST), Los Alamos National Laboratory, the Max-Planck-Institute for Astronomy (MPIA), the Max-Planck-Institute for Astrophysics (MPA), New Mexico State University, Ohio State University, University of Pittsburgh, University of Portsmouth, Princeton University, the United States Naval Observatory, and the University of Washington. 

Funding for SDSS-III has been provided by the Alfred P. Sloan Foundation, the Participating Institutions, the National Science Foundation, and the U.S. Department of Energy Office of Science. The SDSS-III web site is http://www.sdss3.org/. SDSS-III is managed by the Astrophysical Research Consortium for the Participating Institutions of the SDSS-III Collaboration including the University of Arizona, the Brazilian Participation Group, Brookhaven National Laboratory, University of Cambridge, Carnegie Mellon University, University of Florida, the French Participation Group, the German Participation Group, Harvard University, the Instituto de Astrofisica de Canarias, the Michigan State/Notre Dame/JINA Participation Group, Johns Hopkins University, Lawrence Berkeley National Laboratory, Max Planck Institute for Astrophysics, Max Planck Institute for Extraterrestrial Physics, New Mexico State University, New York University, Ohio State University, Pennsylvania State University, University of Portsmouth, Princeton University, the Spanish Participation Group, University of Tokyo, University of Utah, Vanderbilt University, University of Virginia, University of Washington, and Yale University. 

\end{acknowledgements}

\begin{appendix}

\begin{figure*}[t!]
\begin{center}
	\resizebox{14cm}{!}{\includegraphics{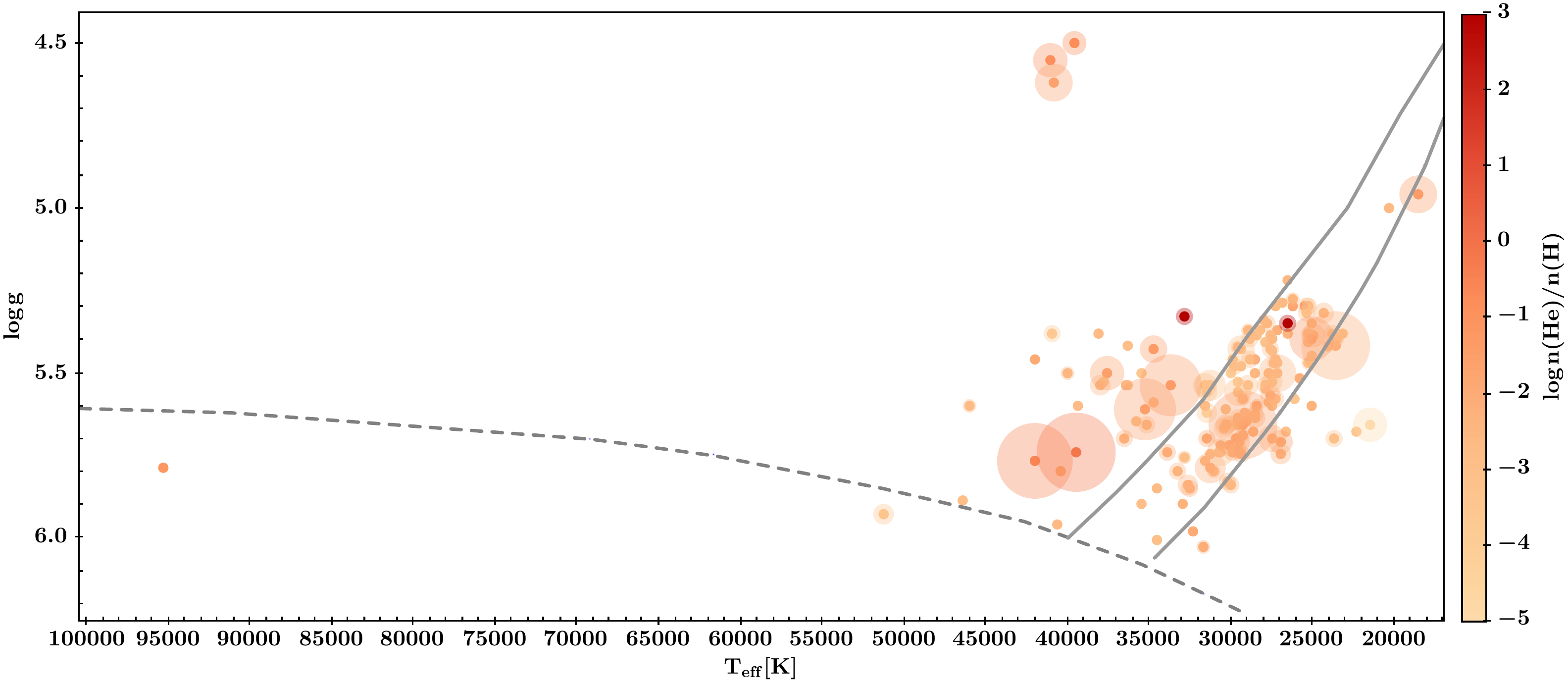}}
	\resizebox{14cm}{!}{\includegraphics{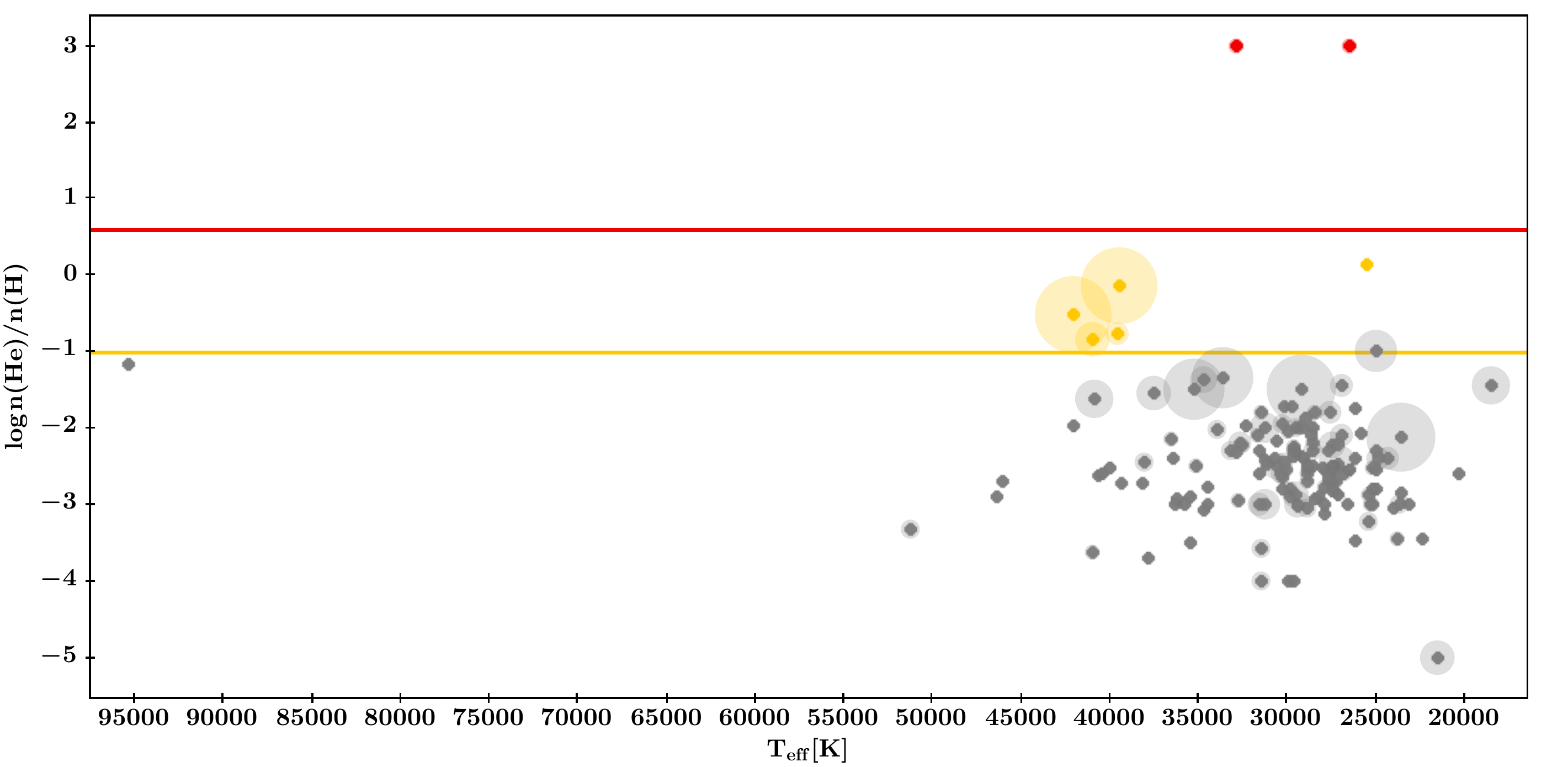}}
\end{center} 
\caption{Upper panel: $T_{\rm eff}-\log{g}$ diagram of the sample of solved close hot subdwarf binaries similar to Fig.~\ref{tefflogg_all}. Lower panel: $T_{\rm eff}-\log{n({\rm He})/n({\rm H})}$ diagram of the solved systems similar to Fig.~\ref{tefflogy_all}. The scaling of the symbol sizes is different and was chosen for better visualisation.}
\label{tefflogg_lit}
\end{figure*}

\section{Previous radial velocity variability studies}\label{rvvar}

Since the known companions in close hot subdwarf binaries are either compact objects like WDs, or very cool M-dwarfs and brown dwarfs, they reveal themselves by the Doppler shifts they induce in the spectral lines of the sdO/B primaries. RV variations are therefore one way to detect close binary hot subdwarfs.

The high close binary fraction of sdB stars was initially discovered by Maxted et al. (\cite{maxted01}). They observed 36 EHB stars in total, five of them being composite sdB binaries and 5 more evolved post-EHB stars. The stars were observed over a timespan of 11 nights in total with the single observing epochs usually separated by one or a few days. The accuracy of their RV measurements was $\sim2-5\,{\rm km\,s^{-1}}$ and they detected significant RV shifts for 21 EHB stars ($58\%$) and one post-EHB star ($20\%$). Excluding the non-variable composite binaries from the counting, the fraction of RV-variable single-lined EHB sdBs was $68\%$. Assuming a certain period distribution and estimating the detection efficiency, Maxted et al. (\cite{maxted01}) concluded that the close binary fraction must be $69\pm9\%$ or more. 

Napiwotzki et al. (\cite{napiwotzki04a}) observed 46 single-lined sdB stars in the course of the SPY survey and found 18 ($39\%$) of them to be RV variable. Two spectra of each star were taken at random epochs over a few years. The RV accuracy was $\sim2\,{\rm km\,s^{-1}}$. Excluding three post-EHB stars with a binary fraction of $67\%$ (two out of three) from the sample, the binary fraction on the EHB was only $37\%$. Also corrections similar to the ones performed by Maxted et al. (\cite{maxted01}) using an updated period distribution only raised the EHB binary fraction to $40\%$, significantly different from the one obtained by Maxted et al. (\cite{maxted01}). Since the sample of Napiwotzki et al. (\cite{napiwotzki04a}) reached down to fainter magnitudes, the authors proposed that a larger fraction of stars from the thick disk or the halo population were observed and that the binary properties of the Galactic populations are different. 

Napiwotzki et al. (\cite{napiwotzki04a}) also determined the RV-variable fraction of helium-rich sdO stars. Out of 23 single-lined He-sdOs they only found one star ($4\%$) to be variable ($5\%$ if corrected as described in Maxted et al. \cite{maxted01}). This provided the first observational hint that the He-sdOs might not be evolutionary related to the sdBs. 

Morales-Rueda et al. (\cite{morales03}) and Copperwheat et al. (\cite{copperwheat11}) extended radial-velocity study of Maxted et al. (\cite{maxted01}). They obtained multi-epoch RVs of 159 sdBs (two of them in the post-EHB stage), derived binary parameters of 51 close binaries and detected significant RV variations for another 20 stars. The accuracies of their RV measurements are quite inhomogeneous ranging from $2$ to more than $20\,{\rm km\,s^{-1}}$. Most of the observed stars have RV epochs with timebases of days, which is much longer than the typical orbital periods. However, several stars have been observed within one night only introducing a bias in favor of very short orbital periods. The authors detected significant RV shifts for 64 of the 125 single-lined EHB stars in this sample ($51\%$), while the two post-EHB stars are both close binaries ($100\%$). In addition 32 objects were found to be composite systems and five of them ($16\%$) to be RV-variable. Copperwheat et al. (\cite{copperwheat11}) did not perform simulations, but estimated the true close binary fraction to range from $46\%$ to $56\%$.

Green et al. (\cite{green08}) showed preliminary results from a study of 407 sdO/Bs as part of a larger survey of hot stars. Using multi-epoch, low-resolution ($9\,{\rm \AA}$), but high-S/N spectra they reported an RV accuracy of $16\,{\rm km\,s^{-1}}$ and calculated the standard deviation of the single RV measurements. In contrast to Napiwotzki et al. (\cite{napiwotzki04a}) they did not detect any significant difference between He-rich and He-poor sdO/Bs in RV variability, which means that variations have been reported in a significant fraction of He-sdO/Bs in their sample.

Kawka et al. (\cite{kawka15}) studied 38 hot subdwarfs randomly selected from a sample of UV-bright objects (Vennes et al. \cite{vennes11}; Nemeth et al. \cite{nemeth12}). The RV follow-up was performed with several different instruments and the overall accuracy was estimated to be smaller than $10\,{\rm km\,s^{-1}}$. Almost all stars have been observed multiple times at random epochs without any bias against longer period systems. 
Six composite systems were detected by visual inspection of the spectra and by fitting their spectral energy distributions, which all did not show significant RV variations ($0\%$). Three stars turned out to be RV-constant He-sdOBs ($0\%$). The sample also included an RV-variable low-mass He-WD progenitor ($100\%$) and a post-EHB sdOB also exhibiting RV variations ($100\%$). From the remaining 27 EHB stars 9 ($33\%$) were found be RV variable. 

Finally, Geier et al. (\cite{geier15b,geier17a}) reported the discovery of 76 new significantly RV-variable hot subdwarfs and 53 candidates in the course of the MUCHFUSS project. However, since this survey aimed at finding short-period binaries with high RV-amplitudes, the target selection was strongly biased and no binary fraction could be derived from this sample. In contrast to other studies, Geier et al. (\cite{geier15b,geier17a}) also targeted 29 He-rich sdOs and sdOBs and discovered six of them to be significantly RV variable. Studying some of those objects in more detail, the authors were not able to determine their orbital parameters and concluded that the irregular RV-variability might be caused by processes other than the presence of close companions. Geier et al. (\cite{geier15b,geier17a}) also reported a mismatch between the RV-variability fraction of sdBs and sdOBs, which should be similar if sdBs are progenitors of sdOBs.

In summary, the previous studies of RV variability in samples of hot subdwarfs found that He-sdO/Bs and sdBs with composite spectra usually do not show significant RV variability, while the variability fractions of EHB and post-EHB stars were rather inconsistent. Selection effects were proposed as possible reason for these inconsistencies. 

\onecolumn

\begin{landscape}
\begin{longtable}{llllllllll}
\caption{\label{data} RV variabilty and spectroscopic parameters for the full sample. $^{a}$Helium abundances without uncertainty are upper (if negative) or lower limits (if positive).} \\
\hline
\noalign{\smallskip}
Name                       &   Class        & ep & $RV_{\rm average}$   & $\Delta RV_{\rm max}$ & $\log{p}$    &  $T_{\rm eff}$        & $\log{g}$        & $\log{y}$$^{a}$     & Reference  \\
                           &                &        & ${\rm [km\,s^{-1}]}$ & ${\rm [km\,s^{-1}]}$  &              &  [K]                 &                  &                  &            \\
\noalign{\smallskip}
\hline
\noalign{\smallskip}
PG2358+107                 &   sdB          &   2    & 31   $\pm$   4     &     9  $\pm$ 10     &        -0.434    &  27100  $\pm$ 430    & 5.54  $\pm$ 0.07 & -2.72$\pm$ 0.11  &   2016ApJ...818..202L  \\    
PG2359+197                 &   sdB+WD       &   2    & -135 $\pm$   4     &     14 $\pm$ 8      &        -1.0478   &  31434  $\pm$ 500    & 5.56  $\pm$ 0.05 & -3.58$\pm$ 0.05  &   2005A\&A...430..223L  \\
SDSSJ000307.06+241211.6    &   sdB          &   9    & 0    $\pm$   4     &     35 $\pm$ 19     &        -0.1199   &  27448  $\pm$ 481    & 5.52  $\pm$ 0.09 & -2.72$\pm$ 0.43  &                        \\
SDSSJ000607.88-010320.8    &   He-sdO       &   6    & -242 $\pm$   5     &     48 $\pm$ 21     &        -0.7693   &  45456  $\pm$ 666    & 5.87  $\pm$ 0.07 & 3.62 $\pm$ 0.72  &                        \\
PG0009+191                 &   sdB          &   2    & -54  $\pm$   4     &     73 $\pm$ 9      &        -13.1608  &  30569  $\pm$ 296    & 5.89  $\pm$ 0.06 & -2.62$\pm$ 0.15  &   2019ApJ...881....7L  \\
\hline\hline
\end{longtable}
\end{landscape}

\end{appendix}

\end{document}